\documentclass[12pt,preprint]{aastex}

\shorttitle{Magnetized Radiation-Dominated Disks}
\shortauthors{Turner et al.}

\newcommand{\divv}{{\bf\nabla\cdot v}}
\newcommand{\ses}{\sigma_{es}}

\begin{document}
\title{Local Three-Dimensional Simulations of Magneto-Rotational
Instability in Radiation-Dominated Accretion Disks}

\author{N. J. Turner\altaffilmark{1,2}, J. M. Stone\altaffilmark{1,3},
J. H. Krolik\altaffilmark{4}, \& T. Sano\altaffilmark{1,3,5}}

\altaffiltext{1}{Astronomy Department, University of Maryland, College
Park MD 20742, USA}

\altaffiltext{2}{Physics Department, University of California, Santa
Barbara CA 93106, USA}

\altaffiltext{3}{Department of Applied Mathematics and Theoretical
Physics, Centre for Mathematical Sciences, Wilberforce Road, Cambridge
CB3 0WA, UK}

\altaffiltext{4}{Department of Physics and Astronomy, Johns Hopkins
University, Baltimore, MD 21218, USA}

\altaffiltext{5}{Institute of Laser Engineering, Osaka University,
Suita, Osaka 565-0871, Japan}

\begin{abstract}
We examine the small-scale dynamics of black hole accretion disks in
which radiation pressure exceeds gas pressure.  Local patches of disk
are modeled by numerically integrating the equations of radiation MHD
in the flux-limited diffusion approximation.  The shearing-box
approximation is used, and the vertical component of gravity is
neglected.  Magneto-rotational instability (MRI) leads to turbulence
in which accretion stresses are due primarily to magnetic torques.
When radiation is locked to gas over the length and time scales of
fluctuations in the turbulence, the accretion stress, density
contrast, and dissipation differ little from those in the
corresponding calculations with radiation replaced by extra gas
pressure.  However, when radiation diffuses each orbit a distance that
is comparable to the RMS vertical wavelength of the MRI, radiation
pressure is less effective in resisting squeezing.  Large density
fluctuations occur, and radiation damping of compressive motions
converts $P dV$ work into photon energy.  The accretion stress in
calculations having a net vertical magnetic field is found to be
independent of opacity over the range explored, and approximately
proportional to the square of the net field.  In calculations with
zero net magnetic flux, the accretion stress depends on the portion of
the total pressure that is effective in resisting compression.  The
stress is lower when radiation diffuses rapidly with respect to the
gas.  We show that radiation-supported Shakura-Sunyaev disks accreting
via internal magnetic stresses are likely in their interiors to have
radiation marginally coupled to turbulent gas motions.
\end{abstract}


%%%%%%%%%%%%%%%%%%%%%%%%%%%%%%%%%%%%%%%%%%%%%%%%%%%%%%%%%%%%%%%%%%%%%%%%%%%%%%%
\section{INTRODUCTION}

Owing to the difficulty of removing angular momentum from infalling
gas, material accreting on a black hole likely first accumulates in a
disk supported against radial gravity by its rotation.  The evolution
of the disk is governed by the extraction of its angular momentum, and
the fate of the released gravitational energy.  A possible structure
for the flow was found by \cite{ss73}.  They assumed angular momentum
was transferred outwards within the disk by an effective viscosity of
unknown origin, proportional to the vertically-averaged pressure at
each radius.  The released energy was converted to heat by the same
viscosity, and the disk was cooled by vertical diffusion of photons to
the surfaces.  The disk was supposed to be time-steady, axisymmetric,
and in vertical hydrostatic balance.

In the inner regions of Shakura-Sunyaev models with luminosities near
the Eddington limit, radiation pressure is much larger than gas
pressure, and provides the main means of support in the vertical
direction.  If the stress which transports angular momentum is
proportional to total pressure, the radiation-dominated regions are
viscously \citep{le74}, thermally \citep{ss76}, and convectively
\citep{bkb77} unstable.  These instabilities might prevent the
formation of a steady disk.  However, if the effective viscosity
results from magnetic activity, buoyancy of the field may limit the
stress to a value proportional to gas pressure alone, resulting in a
thermally and viscously stable configuration \citep{sc89}.  The
structure of the inner parts of accretion flows onto black holes
remains unknown.

A physical mechanism for transfer of angular momentum through disk gas
is magneto-rotational instability (MRI) \citep{bh91}.  In this paper
we examine the effects of MRI in radiation-dominated accretion disks.
The MRI is a local linear instability, driven by exchange of angular
momentum along magnetic field lines linking material at different
distances from the black hole.  Its fastest mode is axisymmetric and
grows at three-quarters the orbital angular frequency $\Omega$.  The
wavelength of fastest growth is fixed by a balance between Coriolis
and magnetic tension forces, and is approximately the distance $2\pi
v_{A}/\Omega$ that Alfv\'en waves travel in an orbital period
\citep{bh98}.

In local three-dimensional MHD simulations without radiation, MRI
leads to turbulence in which magnetic and hydrodynamic stresses
transport angular momentum outwards \citep{hgb95}.  The magnitudes of
the stresses depend on the geometry of the magnetic field.  If the
field has a net vertical flux, stresses are large and depend on the
net flux.  If the field instead has a net azimuthal flux, the stresses
are weaker for the same pressure in the net component.  If the field
has zero net flux, stresses are weaker still, and after a few tens of
orbits, are independent of initial field strength \citep{hgb96}.
Which of these magnetic configurations is most appropriate for disks
around black holes may depend on non-local effects of outflow or
buoyancy, which could lead to a build-up of net magnetic flux in
accreting material.

The effects of MRI in radiation-dominated disks are uncertain.  The
range of linearly-unstable wavelengths is unaffected by photon
diffusion.  However, when azimuthal magnetic pressure exceeds gas
pressure, and photons diffuse more than a vertical MRI wavelength
$\lambda_z=2\pi v_{Az}/\Omega$ per orbit, the growth rate of the
axisymmetric MRI is reduced by a factor roughly $v_{A\phi}/c_g$, where
$v_{Az}$ and $v_{A\phi}$ are the vertical and azimuthal Alfv\'en
speeds, and $c_g$ the gas acoustic speed \citep{bs01}.  Linear growth
may be slow when magnetic pressure is greater than gas pressure, but
much less than total pressure.

The MRI converts gravitational energy into magnetic and kinetic
energy.  Dissipation of the magnetic fields may heat the gas.  Part of
the kinetic energy may be converted directly to photon energy, by
radiative damping of compressive disturbances having wavelengths
shorter than the distance photons diffuse per wave period
\citep{ak98}.  In local axisymmetric MHD simulations, the MRI drives
turbulence with density contrasts as great as the ratio of magnetic to
gas pressure.  Escape of radiation from the compressed regions damps
the motions [\cite{tss02}, hereafter TSS].  These results suggest that
magnetized turbulence may be important for heating as well as angular
momentum transfer in radiation-dominated accretion disks.

Here we extend the radiation MHD calculations of TSS to three
dimensions, neglecting stratification.  In three-dimensional
calculations, the turbulence may reach a time-averaged steady state
lasting many orbital periods.  We examine the effects of radiation
diffusion on the regeneration of magnetic field, the accretion
stresses, and the damping of the turbulence.

%%%%%%%%%%%%%%%%%%%%%%%%%%%%%%%%%%%%%%%%%%%%%%%%%%%%%%%%%%%%%%%%%%%%%%%%%%%%%%%
\section{DOMAIN, EQUATIONS SOLVED, AND NUMERICAL METHODS
\label{sec:method}}

We use the local shearing box approximation \citep{hgb95}.  The domain
is a small patch of the disk, centered at the midplane a distance
$R_0$ from the axis of rotation.  Curvature along the direction of
orbital motion is neglected, and the patch is represented in Cartesian
coordinates co-rotating at the Keplerian orbital frequency $\Omega_0$
for domain center.  Coriolis and tidal forces in the rotating frame
are included, while the component of gravity perpendicular to the
midplane is neglected.  The local coordinates $(x, y, z)$ correspond
to distance from the origin along the radial, orbital, and vertical
directions, respectively.  The azimuthal and vertical boundaries are
periodic, and the radial boundaries are shearing-periodic.  Fluid
passing through one radial boundary appears on the other at an azimuth
which varies in time according to the difference in orbital speed
across the box.  The difference is computed using a Keplerian profile
linearised about domain center.  Owing to the periodic boundary
conditions, net magnetic flux is expected to be constant over time,
provided the net radial flux is initially zero.

The equations of radiation magnetohydrodynamics are solved correct to
zeroth order in $v/c$.  Relativistic effects are neglected, the
flux-limited diffusion (FLD) approximation is used, and gas and
radiation are assumed to be in LTE at separate temperatures.  The
equations are
\begin{equation}\label{eqn:cty}
{D\rho\over D t}+\rho\divv=0,
\end{equation}
\begin{equation}\label{eqn:gasmomentum}
\rho{D{\bf v}\over D t} = -{\bf\nabla}p
	+ {1\over 4\pi}({\bf\nabla\times B}){\bf\times B}
        + {\chi\rho\over c}{\bf F} - 2\rho{\bf\Omega_0\times v} +
	3\rho\Omega_0^2 x{\bf\hat x},
\end{equation}
\begin{equation}\label{eqn:radenergy}
\rho{D\over D t}\left({E\over\rho}\right) =
	- {\bf\nabla\cdot F} - {\bf\nabla v}:\mathrm{P}
	+ \kappa\rho(4\pi B - c E),
\end{equation}
\begin{equation}\label{eqn:gasenergy}
\rho{D\over D t}\left({e\over\rho}\right) =
	- p\divv - \kappa\rho(4\pi B - c E),
\end{equation}
\begin{equation}\label{eqn:radmomentum}
{\bf F} = -{c\Lambda\over\chi\rho}{\bf\nabla}E,
\end{equation}
and
\begin{equation}\label{eqn:induction}
{\partial{\bf B}\over\partial t} = {\bf\nabla\times}({\bf v\times B})
\end{equation}
\citep{mm84,smn92,hgb95}.  Here $\rho$, ${\bf v}$, $e$, and $p$ are
the gas density, velocity, internal energy density, and pressure,
respectively, and ${\bf B}$ is the magnetic field.  In the Coriolis
term in the equation of motion~\ref{eqn:gasmomentum}, ${\bf\Omega_0}$
is the orbital angular frequency at domain center.  Its direction is
parallel to the rotation axis.  In the tidal term, ${\bf\hat x}$ is a
unit vector along the radial direction.  The photons are represented
by their frequency-integrated energy density $E$, energy flux ${\bf
F}$, and pressure tensor $\rm P$.  Total opacity $\chi$ is the sum of
electron scattering opacity $\sigma=0.4$ cm$^2$g$^{-1}$ and free-free
absorption opacity $\kappa=10^{52} \rho^{9/2} e^{-7/2}$
cm$^2$g$^{-1}$.  In some calculations, scattering opacities higher or
lower than the electron scattering value are used.  The gas cools by
emitting photons at a rate proportional to the blackbody value
$B=\sigma_B T_g^4/\pi$, where $\sigma_B$ is the Boltzmann constant,
$T_g=p\mu/({\cal R}\rho)$ the gas temperature, $\mu=0.6$ the
dimensionless mean mass per particle, and ${\cal R}$ the gas constant.
In equation~\ref{eqn:radmomentum}, the flux-limiter $\Lambda$ is equal
to $1/3$ in optically-thick regions.  Causality is preserved in
regions where radiation energy density varies over optical depths less
than unity, by reducing the limiter towards zero \citep{lp81}.  The
equations are closed by assuming an ideal gas equation of state
$p=(\gamma-1)e$, with $\gamma=5/3$.  Shocks are captured using an
artificial viscosity in regions of convergence according to the
standard prescription of \cite{vr50}.  We solve
equations~\ref{eqn:cty}-\ref{eqn:induction} using a three-dimensional
version of the Zeus code \citep{sn92a,sn92b} with its FLD module
\citep{ts01}.

In test calculations of magnetized turbulence in a shearing box, the
fraction of the work done by shear that is lost from the domain
through numerical effects is as great as 95\% in ideal MHD, and as
little as 20\% when an Ohmic resistivity is included.  We infer that
the main numerical energy losses in the ideal-MHD calculations may
occur in treating the magnetic terms in
equations~\ref{eqn:gasmomentum} and~\ref{eqn:induction}.  For two of
the calculations described in section~\ref{sec:bz}, the internal
energy scheme usually used in Zeus to solve
equation~\ref{eqn:gasenergy} is therefore replaced by a partial total
energy scheme.  This is intended to capture as heat the energy that
would otherwise be lost through numerical dissipation of magnetic
fields.  During the magnetic part of each timestep, total energy is
conserved.  Immediately before the magnetic fields are updated, the
gas internal energy density $e$ in each zone is replaced by the sum
$e_T$ of gas internal, magnetic, and kinetic energy densities.  The
field is updated as usual, with time-centered EMFs computed by the
Method of Characteristics \citep{hs95}.  The same EMFs are used to
find the Poynting fluxes ${\bf S}=-{1\over 4\pi}{\bf (v\times B)\times
B}$ of electromagnetic energy across zone faces.  Total energy is
moved from zone to zone according to these fluxes, using the
difference form of
\begin{equation}
{\partial e_T \over\partial t} = - {\bf\nabla\cdot S}.
\end{equation}
Accelerations due to Lorentz forces are applied to the velocities in
the usual way.  Finally, the new magnetic and kinetic energies are
subtracted from the new total energies, and the remainder is assigned
to gas internal energy.

Results using the internal and partial total energy schemes were
compared against a one-dimensional analytic solution for propagation
of non-linear torsional Alfv\'en waves in a uniform fluid
\citep{sa98}.  The waves have moving nulls in the components of the
field transverse to the direction of propagation.  At the nulls,
numerical losses of magnetic field may be rapid.  Tests were carried
out with longitudinal and transverse fields initially equal.  The gas
was either stationary on the grid, or moving in the direction of wave
propagation at nine times the Alfv\'en speed.  When gas and magnetic
pressures were equal, results using the two schemes were similar.
When gas pressure was 1\% of magnetic pressure, numerical losses of
field near magnetic nulls in the internal energy scheme led to
longitudinal total pressure variations.  Longitudinal motions grossly
distorted the wave within a few oscillations.  With the partial total
energy scheme, total pressure was almost constant across the nulls,
and the wave shape changed only slightly over ten oscillations.  For
the case of stationary fluid, total energy decreased less than one
part in $10^5$ over the same period.  In additional tests, results for
the \cite{bw88} MHD Riemann problem differed little between the two
energy schemes.

%%%%%%%%%%%%%%%%%%%%%%%%%%%%%%%%%%%%%%%%%%%%%%%%%%%%%%%%%%%%%%%%%%%%%%%%%%%%%%%
\section{INITIAL CONDITIONS
\label{sec:ic}}

Initial conditions are selected from a \cite{ss73} model with
parameters appropriate for an active galactic nucleus.  The central
mass $M=10^8 M_\odot$, luminous efficiency $\eta=L/({\dot M}c^2)=0.1$,
and accretion rate $\dot M$ is 10\% of the Eddington value ${\dot
M}_E=2.65\times 10^{-9} (M/M_\odot) \eta^{-1}$ $M_\odot$~yr$^{-1}$.
In choosing the initial state only, the ratio of stress to total
pressure is set to $\alpha=0.01$.  The domain is uniformly filled with
gas having the density and midplane temperature of the Shakura-Sunyaev
model at the central radius $R_0$.  Radiation energy density is chosen
for thermal equilibrium with the gas.  The calculations are centered
either at location~I, where $R_0=67.8 r_G$ and radiation pressure is
125 times gas pressure, or at location~II, where $R_0=177 r_G$ and
radiation pressure is 10 times gas pressure.  The gravitational radius
$r_G = G M/c^2$.  Location~I is identical to location~A discussed by
TSS, while location~II is considered in section~\ref{sec:zn} because
the vertical MRI wavelength may be unresolved in calculations with
zero net magnetic flux at location~I.  Conditions at the two locations
are listed in table~\ref{tab:ic}.  At these radii, the Shakura-Sunyaev
model has large electron scattering optical depths $\tau_\mathrm{es}$
and large effective free-free optical depths $\tau^*_\mathrm{ff}$.
The timescales for free-free energy exchange between gas and radiation
are much less than the orbital periods, consistent with the assumption
of initial thermal equilibrium.  Additional sources of absorption
opacity likely would have little further effect provided Thomson
scattering remained the largest contribution to the total.

\begin{deluxetable}{rll}
\tablewidth{0pt}
\tablecaption{Initial conditions\label{tab:ic}}
\tablehead{& Location I & Location II }
\startdata
$R_0/r_G$               &$67.8$                 &$177$                  \\
$H/r_G$                 &$1.83$                 &$1.83$                 \\
$c_r/c_g$               &$10$                   &$2.83$                 \\
$P/p$                   &$125$                  &$10$                   \\
$\rho$/g~cm$^{-3}$      &$2.89\times 10^{-9}$   &$1.22\times 10^{-8}$   \\
$T_g=T_r$/K             &$2.71\times 10^5$      &$1.89\times 10^5$      \\
$\tau_\mathrm{es}$      &$6.2\times 10^4$       &$2.6\times 10^5$       \\
$\tau^*_\mathrm{ff}$    &$4.5\times 10^2$       &$7.5\times 10^3$       \\
\enddata
\end{deluxetable}

The height and width of the domain in all the calculations described
here are set to the half-thickness $H$ of the Shakura-Sunyaev model.
The depth along the direction of orbital motion is made four times
greater.  This allows the development of structures extended along the
azimuthal direction, as seen in calculations without radiation effects
\citep{hgb95}.  Photons diffuse from midplane to surface in
approximately $1/\alpha$ orbits in the Shakura-Sunyaev picture.  At
the standard electron scattering opacity, photons diffuse across the
domain height in 50~orbits at both locations~I and~II.

Orbital velocities initially follow the linearised Keplerian profile
required for radial force balance.  Radial and vertical velocities in
each zone are randomly chosen between $-1$\% and $+1$\% of the
radiation acoustic speed.

\subsection{Diffusion of Radiation with Respect to Gas Fluctuations
\label{sec:coupling}}

Since photon diffusion can slow the linear growth of the MRI, it is
possible that the fully-developed turbulence may be affected also.
The growth rate of the fastest axisymmetric linear mode is less than
$\frac{3}{4}\Omega$ when the azimuthal magnetic pressure exceeds the
effective pressure.  The effective pressure is equal to the gas
pressure if radiation is absent, or diffuses quickly.  On the other
hand, the effective pressure is due to gas and radiation together, if
radiation pressure disturbances grow faster than they are erased by
diffusion \citep{bb94,bs01}.  In this section we consider the
conditions under which radiation may provide support against magnetic
forces in turbulence, so that fluctuations in density and radiation
pressure are correlated.

From local MHD simulations with and without radiation, it has been
found that turbulence driven by MRI is anisotropic.  Fluctuations are
thinner on average in the vertical direction than in the radial and
azimuthal [\cite{hgb95,hgb96}; this work, sections~\ref{sec:bz}
and~\ref{sec:zn}].  The characteristic size is the vertical MRI
wavelength $\lambda_z=2\pi v_{Az}/\Omega$, and individual fluctuations
typically last about one orbital period $2\pi/\Omega$.  Photons may be
expected to couple to the turbulence if the average vertical MRI
wavelength is longer than the distance diffused per orbit, $l_D =
\left({c\over 3\chi\rho}{2\pi\over\Omega}\right)^{1/2}$.  This
condition for good coupling may be written
\begin{equation}\label{eqn:coupling}
{B_z^2} > {2c\Omega\over 3\chi},
\end{equation}
or
\begin{equation}
|B_z| > 1.0\times 10^8\, \mathrm{Gauss} \left(M\over
M_\odot\right)^{-1/2} \left(R\over r_G\right)^{-3/4}.
\end{equation}
The criterion is independent of the density provided electron
scattering is the largest opacity.

%%%%%%%%%%%%%%%%%%%%%%%%%%%%%%%%%%%%%%%%%%%%%%%%%%%%%%%%%%%%%%%%%%%%%%%%%%%%%%%
\section{UNIFORM VERTICAL MAGNETIC FIELD
\label{sec:bz}}

In this section we describe results from simulations with initially
uniform, vertical magnetic fields.  The domain is placed at
location~I, where radiation pressure is 125 times gas pressure.  The
grid consists of 32, 128, and 32 zones along the $x$-, $y$-, and
$z$-directions, respectively.

\subsection{Fiducial Calculation
\label{sec:bzfiducial}}

For the fiducial calculation, the initial field strength is chosen so
that the MRI wavelength is one-quarter the domain height, and about
twice the distance photons diffuse per orbital period.  The magnetic
pressure is less than the gas pressure by a factor of five, and less
than the sum of gas and radiation pressures by a factor of 630.

During the first 2.5~orbits of the calculation, the initial random
poloidal velocity perturbations lead to exponential growth of several
linear modes.  As expected, the mode growing fastest has four
wavelengths in the domain height, and is independent of radius.  Its
growth rate is 0.55~times the orbital angular frequency.  When the
radial magnetic field is comparable to the vertical field, inward and
outward moving regions collide and the flow becomes turbulent.
Turbulence continues to the end of the calculation at 50~orbits.  The
other calculations described in this section all pass through the same
stages of linear growth and sustained turbulence.

\begin{figure}
\epsscale{0.65}
\plotone{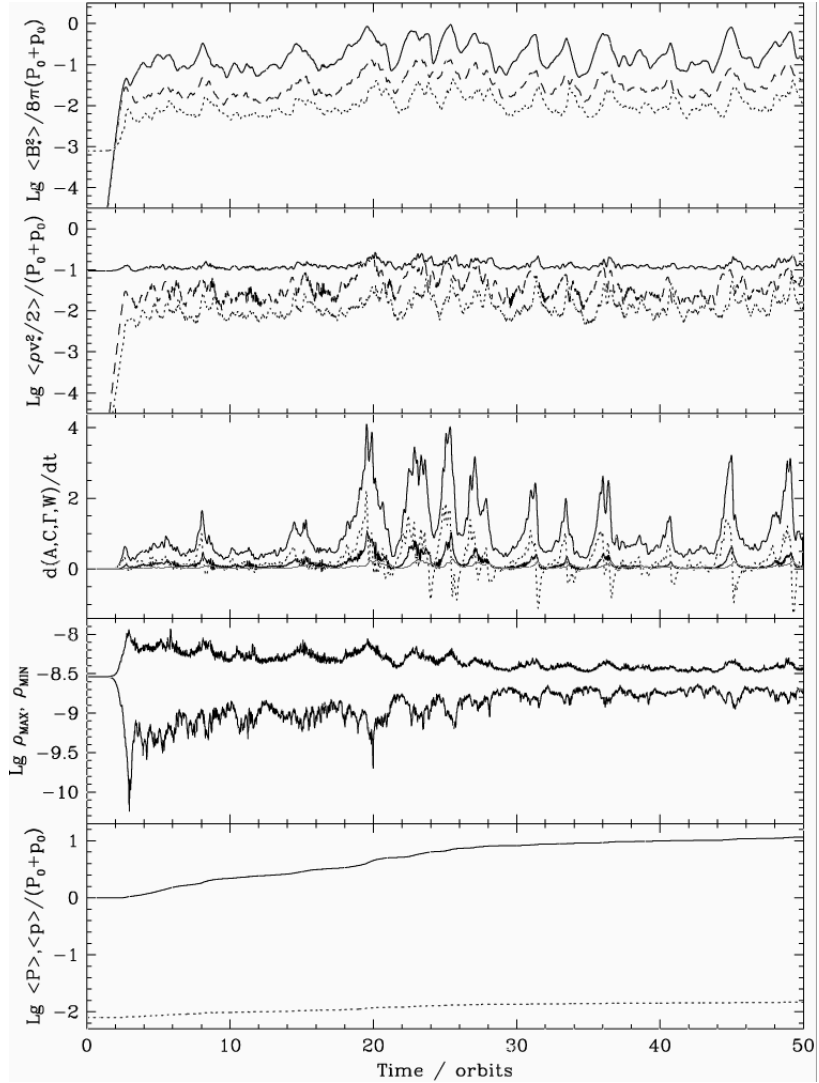}
\caption{Time history of the fiducial simulation with initially
uniform vertical field (\S~\ref{sec:bz}).  Domain-averaged magnetic
and kinetic energy densities measured in the local frame are shown in
the top two panels.  The energies in the $x$-, $y$-, and
$z$-components are indicated by dashed, solid, and dotted lines,
respectively.  The units are the initial radiation pressure $P_0$ plus
gas pressure $p_0$.  Domain-integrated heating rates are plotted in
the third panel, in units of the initial total energy divided by the
orbital period.  Work done by accretion stresses is shown by the upper
solid line, change in total energy by the dotted line, compressive or
$P dV$ heating by the middle solid line, and artificial viscous
heating by the lower, thin solid line.  The largest and smallest
densities in the domain are shown in the fourth panel, and radiation
(solid) and gas (dotted) pressures in the bottom panel.
\label{fig:bztime}}
\end{figure}

The time evolution of the fiducial calculation is shown in
figure~\ref{fig:bztime}.  At the start of the turbulent stage, total
magnetic pressure increases by about two orders of magnitude owing
mostly to growth of the azimuthal field.  Thereafter, magnetic
pressure varies irregularly by an order of magnitude.  Its time
average between 10~and 50~orbits is 59\% of initial gas plus radiation
pressure.  The variations are caused by repeated formation and
disruption of channel flows, similar to those observed in calculations
without radiation by \cite{si01}.  During periods of field growth, the
flow is nearly axisymmetric, and consists of layers of inward and
outward-moving fluid alternating along the vertical direction.  Radial
and azimuthal field strengths increase in lockstep, as in the linear
axisymmetric MRI.  Near peak field strength, the square root of the
domain-averaged squared vertical MRI wavelength may exceed the domain
height.  During periods of field decay, the flow is slower and less
ordered, and the RMS vertical MRI wavelength becomes shorter than the
domain height.  Like the magnetic pressure, the accretion stress has a
well-defined saturated value.  Its time average from 10 to 50 orbits
is 36\% of initial gas plus radiation pressure.  When averaged over
intervals longer than ten orbits, the magnetic pressure and stress
vary little with time.  The mean ratio of the magnetic and
hydrodynamic accretion stresses is 5.3, and angular momentum transfer
is due largely to the magnetic stress.

Gas and radiation remain close to mutual thermal equilibrium
throughout the fiducial calculation due to photon emission and
absorption, while their pressures increase with time.  Over 50~orbits,
radiation pressure increases twelve-fold, gas pressure almost a factor
two.  The increases are due mostly to $P dV$ work done by the flow on
the photons.  Integrated from 10 to 50 orbits and over the domain, the
compressive heating $\langle\langle C\rangle\rangle = -\int dt \int
\left(P+p\right){\bf\nabla\cdot v}\,dx\,dy\,dz$ and artificial viscous
heating $\langle\langle A\rangle\rangle = -\int dt \int {\rm
Q}\,{\bf:\nabla v}\,dx\,dy\,dz$ are 15\% and 6\%, respectively, of the
work $\langle\langle W\rangle\rangle = {3\over 2}\Omega_0H \int dt
\int_X w_{xy}\,dy\,dz$ done by accretion stresses during this period.
Here ${\rm Q}$ is the tensor artificial viscosity, and the subscript
$X$ indicates integration over the radial boundary \citep{hgb95}.
Defining total energy $\Gamma$ as the sum of radiation, internal,
magnetic, kinetic, and gravitational potential energies in the local
frame, the total energy increase between 10~and 50~orbits is just 21\%
of the energy added to the box by accretion stresses.  The remainder
of the work done is removed through numerical losses of magnetic and
kinetic energy.  The density contrast $\rho_{MAX}/\rho_{MIN}$ is more
than an order of magnitude during the first few orbits of turbulence,
but decreases over time, reaching about two at 50~orbits.

Radiation diffusion is an excellent approximation throughout the
calculation.  At all times, the zone with the lowest density has an
optical depth greater than~19.  The components of the Eddington tensor
differ from their isotropic values by less than $10^{-6}$, and
flux-limiting is unimportant.  The coupling between turbulent
fluctuations in gas and radiation varies, owing to the changes in the
strength of the magnetic field, but is generally good.  The ratio of
the vertical MRI wavelength to the diffusion length, computed using
the domain-averaged vertical magnetic pressure, ranges from 4, when
the field is weakest, to 13, when the field is strongest.  Thus,
equation~\ref{eqn:coupling} is satisfied throughout the evolution.
The time-averaged RMS fluctuation in radiation pressure is comparable
to the fluctuation in magnetic pressure, indicating photons are
sufficiently coupled to the turbulence to provide pressure support
against magnetic forces.

The ratio of magnetic accretion stress to vertical magnetic pressure
ranges from 8 to 20, and has time-average 12.5.  The ratio varies in
part due to formation and break-up of channel flows.  Radial and
azimuthal fields grow in the channel flows through MRI, leading to
increases in the ratio of domain-averaged $-B_xB_y/4\pi$ to
$B_z^2/8\pi$.  When the channels are disrupted, some radial and
azimuthal fields are rotated in the turbulence to become vertical, and
the ratio of stress to pressure declines.  From the top panel of
figure~\ref{fig:bztime}, it may be seen that the peaks in vertical
magnetic energy density occur later than the peaks in the other two
components.

The parameters for all the calculations with initially uniform
vertical magnetic fields are shown in table~\ref{tab:bzparms}.  Each
calculation is given a label containing the letter V to indicate that
its magnetic field has a net vertical flux.  The labels of those
including radiation effects start with R.  The labels of those without
the radiation terms start with N.  Each label includes a number
indicating the ratio of the domain height to the initial MRI
wavelength, which fixes the magnetic field strength.  For example, the
fiducial calculation has label RV4.  Suffixes indicate a higher
opacity (h), a lower opacity (l), and a change in the seed used in
generating the random initial poloidal velocities (s).  Initial
radiation, magnetic, and gas pressures are listed relative to the gas
pressure at location~I.

\begin{deluxetable}{llllrcl}
\tablewidth{0pt}
\tablecaption{Parameters for simulations with initially uniform
vertical fields\label{tab:bzparms}}
\tablehead{Label&Name&$H/\lambda_{z,0}$&$\sigma/\ses$&
$P_0 :$&$B_0^2/8\pi$&$:p_0$}
\startdata
NV2.5 & 		   &2.5&\nodata&$  0 :$&$0.513  $&$:126$\\
NV4   & Same total pressure&4  &\nodata&$  0 :$&$0.200  $&$:126$\\
NV8   & 		   &8  &\nodata&$  0 :$&$0.0501 $&$:126$\\
NV12  &			   &12 &\nodata&$  0 :$&$0.0223 $&$:126$\\
NV16  & 		   &16 &\nodata&$  0 :$&$0.0125 $&$:126$\\
\\
RV4h & High opacity        &4  &100    &$125 :$&$0.200  $&$:1$  \\
\\
RV2.5 &                    &2.5&1      &$125 :$&$0.513  $&$:1$  \\
RV3   &                    &3  &1      &$125 :$&$0.356  $&$:1$  \\
RV4\tablenotemark{a}&Fiducial&4&1      &$125 :$&$0.200  $&$:1$  \\
RV5   &                    &5  &1      &$125 :$&$0.128  $&$:1$  \\
RV6   &                    &6  &1      &$125 :$&$0.0891 $&$:1$  \\
RV8   &                    &8  &1      &$125 :$&$0.0501 $&$:1$  \\
RV12  &                    &12 &1      &$125 :$&$0.0223 $&$:1$  \\
RV16  &                    &16 &1      &$125 :$&$0.0125 $&$:1$  \\
\\
RV2.5l&			   &2.5&0.25   &$125 :$&$0.513  $&$:1$  \\
RV4l  & Low opacity        &4  &0.25   &$125 :$&$0.200  $&$:1$  \\
RV4ls &        		   &4  &0.25   &$125 :$&$0.200  $&$:1$  \\
RV8l  &                    &8  &0.25   &$125 :$&$0.0501 $&$:1$  \\
RV12l &                    &12 &0.25   &$125 :$&$0.0223 $&$:1$  \\
RV16l\tablenotemark{a}&    &16 &0.25   &$125 :$&$0.0125 $&$:1$  \\
\enddata
\tablenotetext{a}{Also computed using partial total energy scheme as
described in section~\ref{sec:bzheating}.}
\end{deluxetable}

The results of the vertical field calculations are summarized in
table~\ref{tab:bzresults}.  The domain average of quantity $q$ is
indicated by $\langle q\rangle$, the time and domain average by
$\langle\langle q\rangle\rangle$.  RMS values are computed by the
square root of the domain average of the square.  Time averages in
table~\ref{tab:bzresults} are taken between 10 and 50 orbits after the
start of each run.  The quantities listed are (column~1) the label of
the simulation; (2) the RMS ratio of the vertical MRI wavelength to
the distance photons diffuse per orbit; (3) the ratio of magnetic
stress to vertical magnetic pressure; (4) the total accretion stress
$w_{xy}$, in units of the gas plus radiation pressure at location~I;
(5) the density contrast; and (6) the compression heating
$\langle\langle C\rangle\rangle$ and (7) artificial viscous heating
$\langle\langle A\rangle\rangle$ between 10 and 50 orbits, as
fractions of the work $\langle\langle W\rangle\rangle$ done by
accretion stresses during this period.

\begin{deluxetable}{lccccccc}
\tabletypesize{\scriptsize}
\tablewidth{0pt}
\tablecolumns{7}
\tablecaption{Results from simulations with initially uniform
vertical fields\label{tab:bzresults}}
\tablehead{
\colhead{Label}
 & \colhead{Coupling}
 & \colhead{Field geometry}
 & \colhead{Total stress}
 & \colhead{Density range}
 & \colhead{Compression}
 & \colhead{Artificial} \\
\colhead{}
 & \colhead{$\sqrt{}\langle\langle B_z^2\rangle\rangle$/}
 & \colhead{$\langle\langle -B_xB_y/4\pi\rangle$}
 & \colhead{$\langle\langle w_{xy}\rangle\rangle$}
 & \colhead{$\langle\langle\rho_{MAX}$}
 & \colhead{heating}
 & \colhead{viscous heating} \\
\colhead{}
 & \colhead{$(2c\Omega_0/3\chi)$}
 & \colhead{$/\langle B_z^2/8\pi\rangle\rangle$}
 & \colhead{$/(P_0+p_0)$}
 & \colhead{$/\rho_{MIN}\rangle\rangle$}
 & \colhead{$\langle\langle C\rangle\rangle / \langle\langle W\rangle\rangle$}
 & \colhead{$\langle\langle A\rangle\rangle
            / \langle\langle W\rangle\rangle$}\\
\colhead{(1)} & \colhead{(2)} & \colhead{(3)} &
\colhead{(4)} & \colhead{(5)} & \colhead{(6)} & \colhead{(7)}
}
\startdata
NV2.5	&\nodata &12.6 &0.700	&1.35	&0.0282	 &0.0699 \\
NV4	&\nodata &11.1 &0.291	&1.68	&0.0217	 &0.0757 \\
NV8	&\nodata &9.75 &0.0883	&1.78	&0.00811 &0.0876 \\
NV12	&\nodata &10.9 &0.0473  &1.77	&-0.00479&0.104  \\
NV16	&\nodata &11.3 &0.0357	&1.74	&-0.00531&0.103	 \\
\\	
RV4h	&71.0	 &12.1 &0.358	&2.10	&0.0534	 &0.0738 \\
\\	
RV2.5	&9.26	 &10.8 &0.547	&2.34	&0.114	 &0.0656 \\
RV3	&8.34	 &11.9 &0.487	&2.96	&0.132	 &0.0659 \\
RV4	&6.97	 &12.5 &0.359	&3.64	&0.146	 &0.0648 \\
RV5	&5.33	 &11.0 &0.188	&3.70	&0.147	 &0.0663 \\
RV6	&5.05	 &10.2 &0.158	&4.11	&0.150	 &0.0687 \\
RV8	&3.62	 &10.1 &0.0845	&4.45	&0.160	 &0.0685 \\
RV12	&2.40	 &10.8 &0.0397	&5.14	&0.170	 &0.0674 \\
RV16	&1.83	 &11.3 &0.0249	&5.02	&0.171	 &0.0696 \\
\\	
RV2.5l  &4.84    &12.7 &0.682   &7.80   &0.196   &0.0596 \\
RV4l	&3.39	 &12.3 &0.324	&11.5	&0.226	 &0.0576 \\
RV4ls   &3.07	 &11.2 &0.257   &13.2   &0.229   &0.0564 \\
RV8l    &1.74	 &10.1 &0.0767  &23.9   &0.258   &0.0572 \\
RV12l	&0.990	 &10.3 &0.0258  &20.4   &0.260   &0.0563 \\
RV16l	&0.722	 &11.0 &0.0149	&18.7	&0.262	 &0.0563 \\
\enddata
\tablecomments{Results are averaged between 10 and 50 orbits.}
\end{deluxetable}

\subsection{Magnetic Field and Accretion Stress
\label{sec:bzstress}}

\begin{figure}
\epsscale{0.9}
\plotone{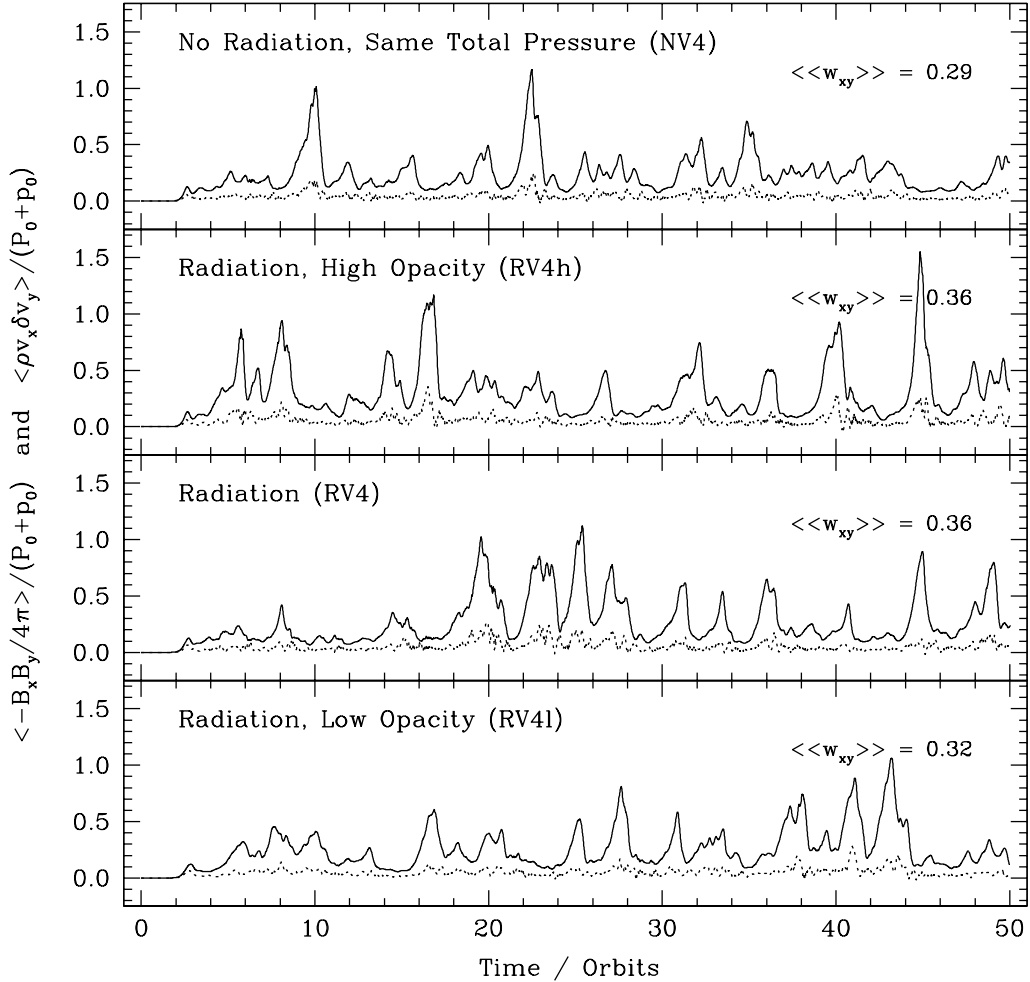}
\caption{Variation of accretion stresses with time in simulations with
initially uniform vertical magnetic fields.  Results are shown from
calculations without radiation (top, NV4), with scattering opacity
100~times the electron scattering value (second panel, RV4h), with
the usual opacities (third panel, RV4), and with scattering opacity
one-quarter of the electron scattering value (bottom, RV4l).  Magnetic
stresses are indicated by solid lines, hydrodynamic stresses by dotted
lines.  In each case, the total accretion stress averaged from 10 to
50 orbits is indicated at upper right.  The stresses are shown in
units of the initial gas plus radiation pressure.
\label{fig:bzstressrad}}
\end{figure}

Over the range of opacities explored here, radiation diffusion has
little effect on the accretion stress (figure~\ref{fig:bzstressrad}).
Results from the fiducial calculation are shown along with those from
versions with scattering opacity increased a hundredfold (RV4h) and
decreased fourfold (RV4l), and radiation pressure replaced by an equal
amount of additional gas pressure (NV4).  The four calculations have
the same initial magnetic field.  The total stresses, averaged over
the domain and over time between~10 and~50 orbits, are 0.36, 0.36,
0.32, and 0.29 times the initial gas plus radiation pressure,
respectively.  The differences in mean stress among the four are much
less than the range of time variation in each.  Stresses in the
calculation which includes gas pressure only are similar to those in
the runs including both gas and radiation.  The time-averaged RMS
ratio of the diffusion scale to the MRI wavelength is 1/71 in the
high-opacity version, 1/7 in the fiducial version, and 1/3.4 in the
low-opacity version.  In these calculations having a net vertical
magnetic flux and strong to marginal coupling between gas and
radiation, the stresses vary little with the diffusion rate.

\begin{figure}
\epsscale{0.65}
\plotone{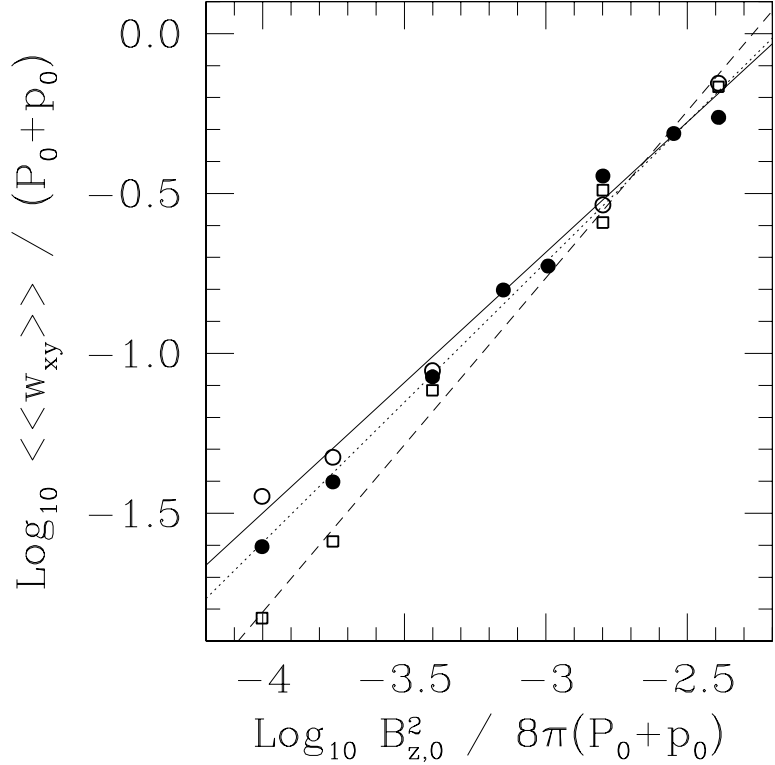}
\caption{Dependence of total accretion stress on initial magnetic
pressure in calculations starting with uniform vertical magnetic
fields.  The stress is averaged over the domain, and from~10 to~50
orbits.  Calculations NV2.5 through NV16 have radiation replaced by
extra gas pressure.  Their results are marked by open circles.  The
best straight-line fit for the results without radiation is shown by a
solid line and has slope $0.82$.  The range of slopes allowed by the
scatter of the points is $\pm 0.04$.  Calculations RV2.5 through RV16
include radiation effects with normal opacities, and are shown by
filled circles.  The best straight-line fit, drawn dotted, has slope
$0.88\pm 0.04$.  Calculations RV2.5l through RV16l have scattering
opacity four times less and are shown by squares.  The best fit line
for the low-opacity results, drawn dashed, has slope $1.04\pm 0.04$.
\label{fig:bzstresspm}}
\end{figure}

The dependence of the mean total accretion stress on initial magnetic
pressure is shown in figure~\ref{fig:bzstresspm}.  The relationship is
approximately linear over the limited range explored, with the stress
about 200~times the initial magnetic pressure.  The uncertainty in the
positions of the points may be gauged by comparing the two open
squares near horizontal position $-2.8$.  These are from RV4l and a
calculation RV4ls identical except that the initial poloidal velocity
perturbations are chosen using a different random number seed.  The
differences between results with and without radiation in
figure~\ref{fig:bzstresspm} are about as large as the differences
resulting from the changed perturbation.

In calculations carried out by \cite{hgb95}, the saturation level is
proportional to the net vertical magnetic field rather than the
vertical magnetic pressure (their figure~6).  The discrepancy may be
due in part to the longer integration time employed here.  Longer
integrations allow better estimates of the time-averaged values of the
fluctuating stresses.  The results in figure~\ref{fig:bzstresspm} are
from runs lasting 50~orbits, whereas the previous calculations lasted
7~to 16~orbits.

The mean ratios of magnetic to hydrodynamic stress in our calculations
are mostly about 5, and the range is from 2.0 to 12.  There is no
clear trend in the ratio of magnetic to hydrodynamic stress with field
strength or opacity.

The time variations in the accretion stress, associated with formation
and breakup of channel flows, decrease in relative amplitude with
decreasing field strength.  In calculations with 12~and 16~wavelengths
initially filling the domain height, no strong channel flows are
observed on the largest scale.  In the weakest-field calculations
NV16, RV16, and RV16l, the MRI wavelength is initially only 2~grid
zones.  However, during the turbulent stage, the field is stronger
than initially owing to squeezing and folding.  The time-averaged RMS
vertical MRI wavelengths in NV16, RV16, and RV16l are 10, 8, and
7~zones, and are probably adequately resolved.  The large stress
variations occur only in the calculations having vertical MRI
wavelength comparable to the box height.

Overall, in the simulations with net vertical flux, the saturated
field strength and accretion stress are similar with and without
radiation effects, whether the photons are strongly or marginally
coupled to the flow.

\subsection{Turbulent Fluctuations
\label{sec:bzfluctuations}}

\begin{figure}
\epsscale{0.5}
\plotone{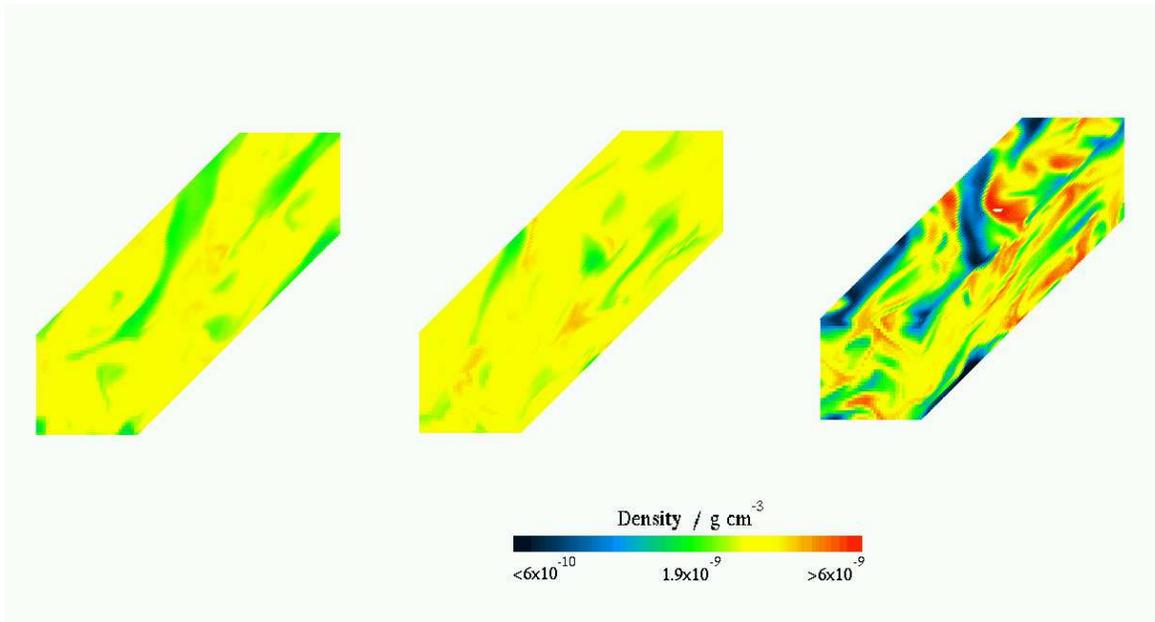}
\caption{Snapshots of the density distribution on the domain faces at
20~orbits, in three calculations with initially uniform vertical
magnetic fields.  The calculation without radiation NV4 is at left,
the high-opacity version RV4h at center, and the fiducial calculation
RV4 at right.  Radius $x$ increases to the right, azimuth $y$ into the
page, and height $z$ upwards.  The common logarithmic density scale
spans one decade, while the density contrast in the fiducial
calculation at this time is a factor $24$.
\label{fig:bzdensity}}
\end{figure}

Large density contrasts may be expected if magnetic pressure is
greater than gas pressure, and radiation diffuses rapidly.  As shown
in figure~\ref{fig:bzdensity} and table~\ref{tab:bzresults}, the
density range is larger in the fiducial calculation RV4 than in the
version NV4 with the radiation replaced by extra gas pressure.  Among
radiation runs with identical initial magnetic fields, the density
contrast is greater in those having weaker coupling of photons to gas.
Run RV4l, where the mean vertical MRI wavelength is 3.4 times the
diffusion length, has a time-averaged density contrast
$\langle\langle\rho_{MAX}/\rho_{MIN}\rangle\rangle=11.5$.  The
fiducial calculation RV4 has a similar mean vertical MRI wavelength, a
diffusion length half as great, and a mean density contrast of~3.6.
In the high-opacity run RV4h, the diffusion length is ten times
shorter again.  The mean density contrast is about~2, and differs
little from that in the run NV4 with radiation replaced by extra gas
pressure.  Among the calculations listed in table~\ref{tab:bzparms},
the time-averaged RMS density fluctuation is well-correlated with the
logarithm of the time-averaged density contrast, indicating that these
two quantities are about equally good measures of the overall density
distribution.

\begin{figure}
\epsscale{0.9}
\plotone{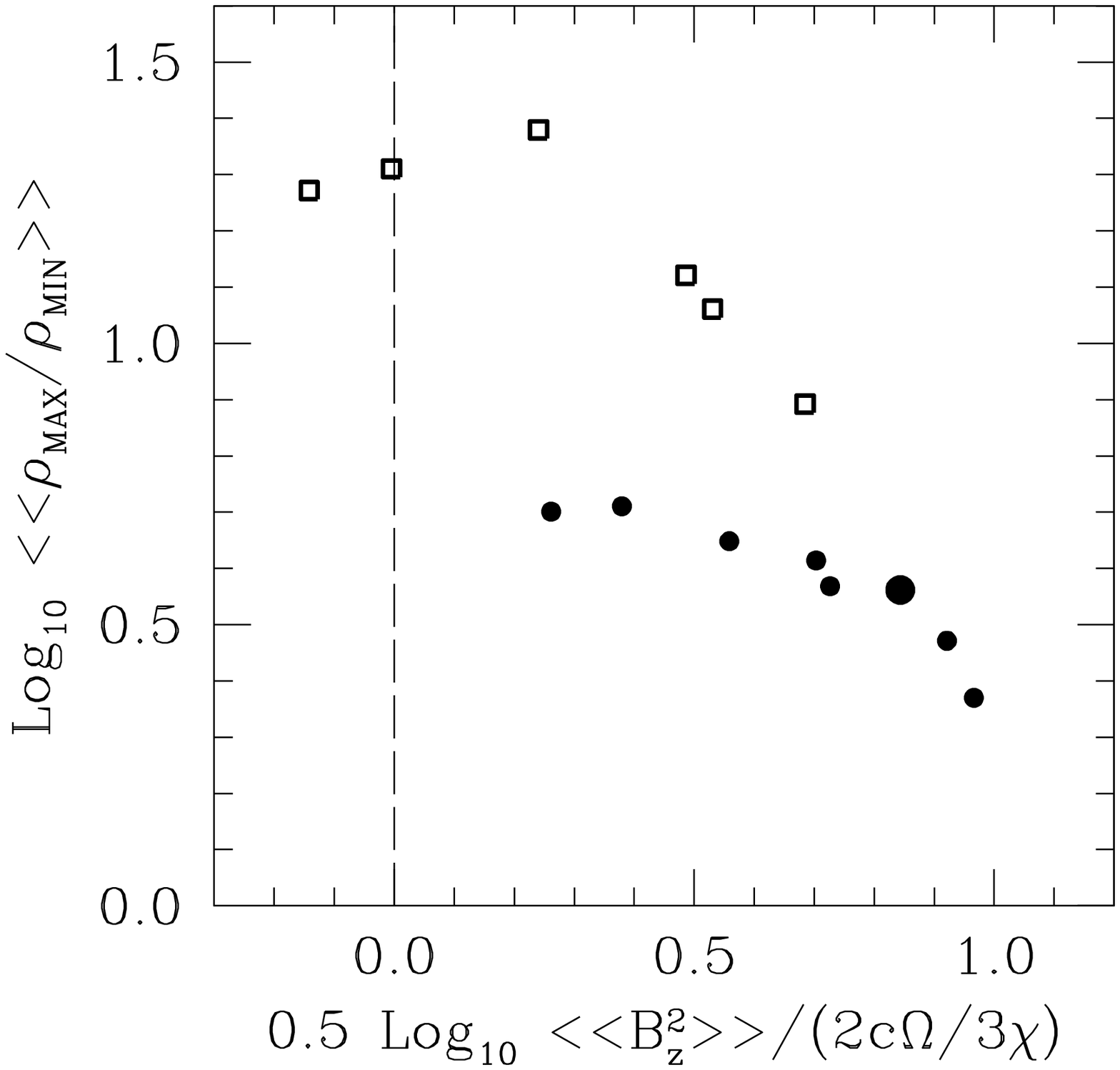}
\caption{Dependence of the mean density contrast on the coupling
between gas and radiation, in calculations with net vertical magnetic
flux.  The horizontal axis is the logarithm of the RMS ratio of the
vertical MRI wavelength to the diffusion scale.  Calculations RV2.5 to
RV16, with standard opacities, are marked by circles.  Those RV2.5l to
RV16l with scattering opacity four times less are marked by squares.
The fiducial calculation RV4 is indicated by a larger circle.  A
dashed line shows where vertical MRI wavelength is equal to diffusion
length.  At a given level of coupling, the density contrast is larger
in the calculation with the smaller opacity and stronger magnetic
field.
\label{fig:bzdrangepmag}}
\end{figure}

Two calculations differing in vertical magnetic pressure can have the
same degree of coupling, provided they differ in opacity in the
inverse proportion.  The squared ratio of the vertical MRI wavelength
to the diffusion scale is proportional to $B_z^2 \chi$, as shown by
equation~\ref{eqn:coupling}.  At a given level of coupling, the
density contrast is found to be greater in the calculation with the
stronger magnetic field (figure~\ref{fig:bzdrangepmag}).  For large
density contrasts, it is necessary that the mean magnetic pressure be
at least comparable to the gas pressure, but not so large that the
vertical MRI wavelength is longer than the radiation diffusion length.
At fixed gas pressure, the maximum density contrast is attained near
the magnetic pressure for which the coupling is marginal.

The density contrast in the fiducial calculation decreases over time
(figure~\ref{fig:bztime}) as the pressure rises.  Gas and radiation
pressures increase steadily owing to dissipation of energy released
through the accretion stresses, whereas magnetic pressure, averaged
over periods of ten orbits, varies little during the turbulent stage.
At 50~orbits, radiation pressure is 12~times larger than initially.
In a weaker-field calculation RV16, the density range near the end of
the turbulent stage is similar to that near the beginning.  The mean
accretion stress here is 14 times lower, and radiation pressure
increases only 1.8-fold over 50~orbits.  The density range depends on
the pressure available to resist squeezing by the magnetic field.

\begin{figure}
\epsscale{0.65}
\plotone{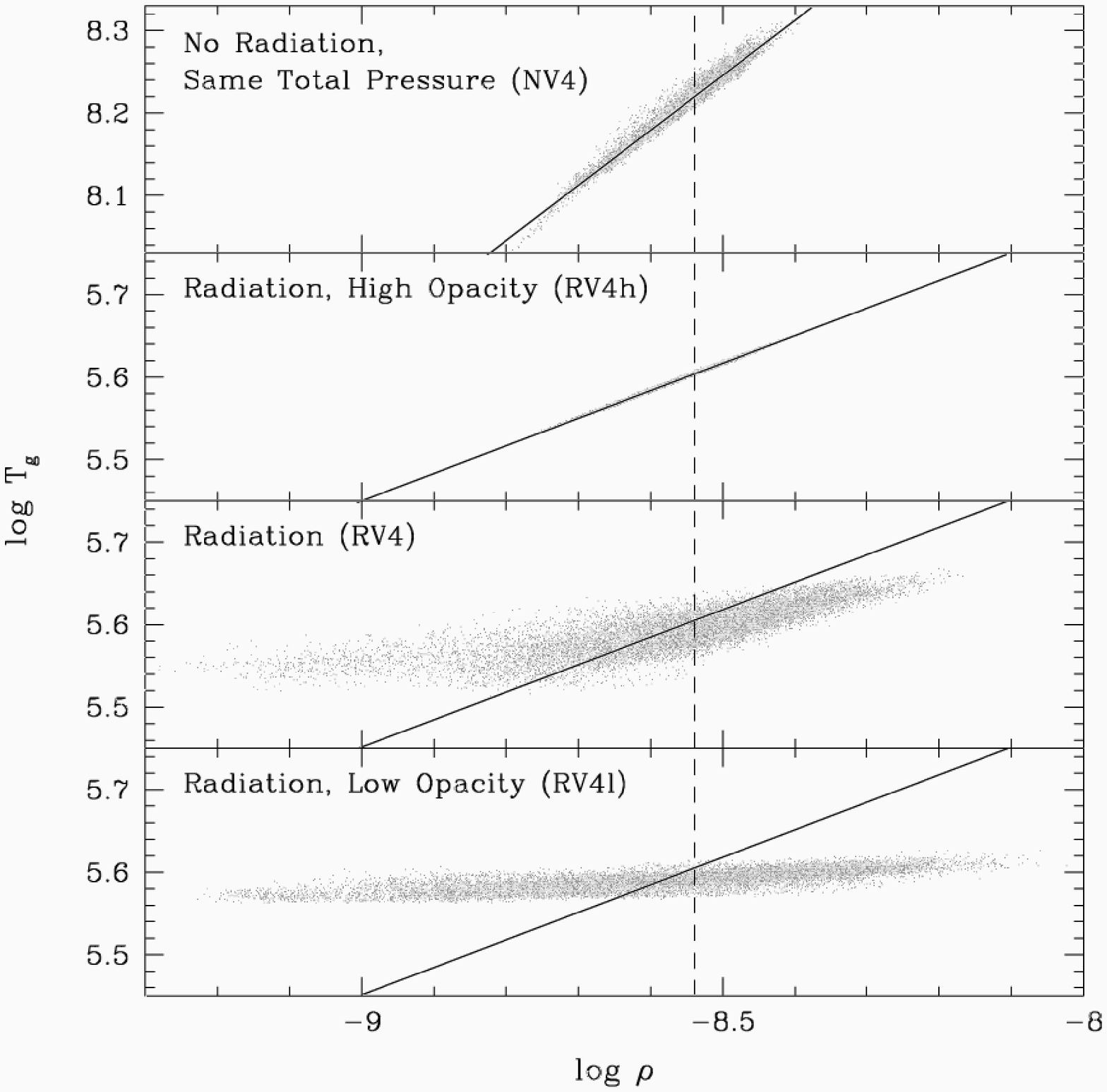}
\caption{Effective gas equations of state in four calculations with
identical, initially uniform vertical magnetic fields.  Gas
temperature and density in every eleventh grid zone are plotted at
20~orbits.  The run without radiation NV4 is shown at top, the
high-opacity run RV4h in the second panel, the fiducial run RV4 third,
and the low-opacity version RV4l at bottom.  Vertical dashed lines
mark the mean density.  The diagonal solid lines have slopes of 2/3
(upper) and 1/3 (remaining panels).
\label{fig:bzstate}}
\end{figure}

The effective equation of state of fluctuations in the turbulence
depends on the degree to which the radiation is coupled to the gas.
When the scattering opacity is large and gas and photons travel
together, the combined fluid has an adiabatic equation of state with
exponent very close to $4/3$ (figure~\ref{fig:bzstate}, second panel).
When the distance photons diffuse per orbit is comparable to the
vertical MRI wavelength, temperature excursions in the gas are damped
by rapid emission or absorption, followed by escape of photons to
other regions.  In this case the gas is close to isothermal
(figure~\ref{fig:bzstate}, bottom panel).  At the highest densities,
radiation is partly trapped, and the temperature rises slightly above
that of the rest of the flow.

\subsection{Heating
\label{sec:bzheating}}

\begin{figure}
\epsscale{0.65}
\plotone{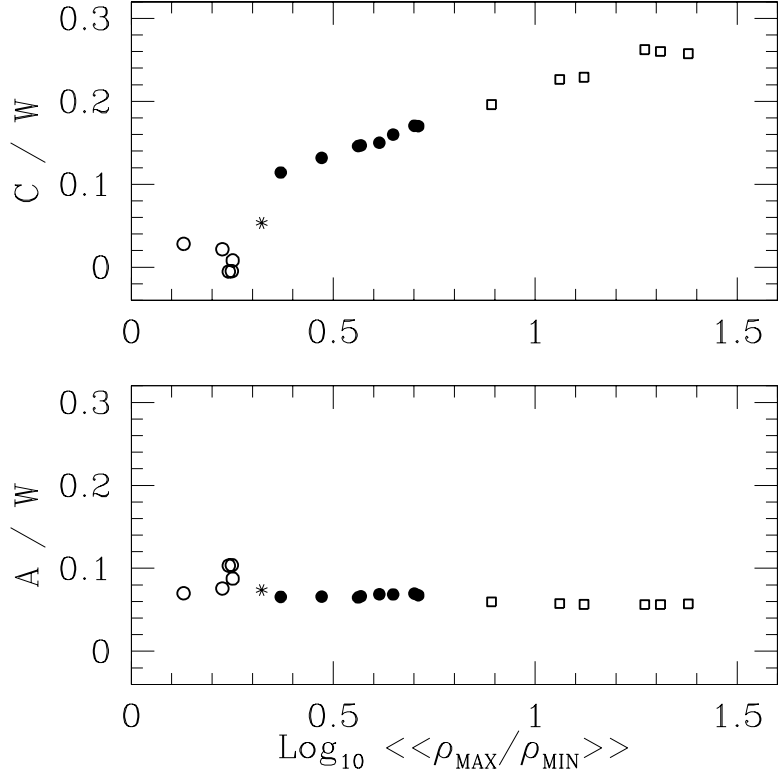}
\caption{Heating due to compression (upper panel) and artificial
viscosity (lower) versus time-averaged density contrast, between
10~and 50~orbits in the calculations listed in
table~\ref{tab:bzparms}.  The vertical axes are labeled in units of
the work done by accretion stresses.  Calculations without radiation
are shown by open circles, with high opacity by stars, standard
opacity by filled circles, and low opacity by squares.  The
compression heating efficiency is well-correlated with density
contrast.  The artificial viscous heating fraction is similar in
calculations with large and small density contrasts.
\label{fig:drangeheating}}
\end{figure}

During the fiducial calculation RV4, the energy density in gas plus
radiation increases by eleven times its initial value.  Heating is
fastest when the magnetic field strength and density contrast are
large (figure~\ref{fig:bztime}).  Three terms in the energy
equations~\ref{eqn:radenergy} and~\ref{eqn:gasenergy} may lead to net
increases in the total internal energy in the domain.  These are the
terms representing compression of radiation and of gas, and artificial
viscous heating of the gas.  Net compression heating may be found in
shocks.  It can also occur when regions of the flow are squeezed, then
cool as radiation diffuses out into the surroundings.  Because
diffusion is thermodynamically irreversible, this mechanism leads to
permanent conversion of $P dV$ work into photon energy.  Work done on
the gas is largely converted into radiation energy within a fraction
of an orbit by emission of photons, and can diffuse away also.  In the
fiducial calculation, the net increase in total internal energy owing
to the two compression terms is 15\% of the work done on the flow by
accretion stresses.  The increase owing to artificial viscous heating
is 6\% of the energy input.  In the calculation NV4 with radiation
replaced by additional gas pressure, diffusion is absent and the
compression heating rate is lower.  The contributions from compression
and artificial viscosity are 2\% and 8\%, respectively.  Among the
calculations listed in table~\ref{tab:bzresults}, the largest ratio of
compression heating to energy input is 26\%, and occurs in the
low-opacity, weak-field run RV16l, where the coupling of photons to
gas is weakest.  The average compression heating fraction varies with
the logarithm of the density contrast in these calculations
(figure~\ref{fig:drangeheating}).  The tight correlation between $P
dV$ heating efficiency and density range indicates that the criteria
for strong compression heating are similar to those for large density
contrast outlined in section~\ref{sec:bzfluctuations}.

The low ratios of the heating rates to the work done in each of the
simulations listed in table~\ref{tab:bzresults} mean that the majority
of the released energy vanishes, and is never deposited in the
internal energy of gas or photons.  In resistive MHD calculations
without radiation, up to 80\% of the work done is dissipated by Ohmic
heating \citep{si01}.  Losses of magnetic field without corresponding
heating may be a major sink of energy in ideal MHD calculations.  Such
losses can occur through numerical diffusion, and through advection of
opposing magnetic fields into a single grid zone.  To examine the
importance of numerical magnetic losses for the energy balance here,
we made a version of the weakest-field low-opacity radiation
calculation RV16l, using the partial total energy scheme described in
section~\ref{sec:method}.  With this scheme, total energy is conserved
during the magnetic field update and Lorentz acceleration portion of
each timestep.  Magnetic energy which disappears during the field
update is placed in the internal energy of the gas.  In the version of
run RV16l using this scheme, the increase in total energy between 10
and 50~orbits is equal to 90.6\% of the work done by the accretion
stresses.  The contributions from compression heating, artificial
viscous heating, and numerical magnetic losses are 23.6, 5.51, and
61.5\% of the work done, respectively.  The majority of the remaining
9.4\% may disappear through losses of kinetic energy in the momentum
transport substep.  The mean total stress, 0.0204~times the initial
gas plus radiation pressure, differs only slightly from that in the
version using the ordinary internal energy scheme.  The sum of gas and
radiation pressures increases 3.4-fold over 50~orbits, whereas in the
version RV16l using the usual internal energy scheme, the increase is
only 60\%.  Gas and radiation remain near thermal equilibrium despite
the additional heating, owing to the emission and absorption of
photons.  In a version of the fiducial run RV4 using the partial total
energy scheme, the extra heating leads to such rapid increases in gas
and radiation pressures that the flow becomes almost incompressible
after a few orbits of turbulence, and $P dV$ heating largely ceases.
These results indicate that numerical losses of magnetic energy must
be considered when using MHD simulations to examine the energy budgets
of accretion disks.

%%%%%%%%%%%%%%%%%%%%%%%%%%%%%%%%%%%%%%%%%%%%%%%%%%%%%%%%%%%%%%%%%%%%%%%%%%%%%%%
\section{FIELDS WITH ZERO NET FLUX
\label{sec:zn}}

Magnetic fields with net vertical flux, such as those used in
section~\ref{sec:bz}, cannot be completely destroyed in the
shearing-box approximation owing to the periodic boundaries.  The mean
pressure in the vertical component of the field cannot fall below its
initial value.  In this section we consider fields with zero flux.
Under these conditions, dissipation and dynamo action may lead to
fields either weaker or stronger than initially.  The starting
magnetic field chosen has strength independent of position, and
direction varying with radius $x$.  Its components are
\begin{equation}\label{eqn:znzbz}
B_z = B_0 \sin {2\pi x \over H},
\end{equation}
\begin{equation}\label{eqn:znzby}
B_y = B_0 \cos {2\pi x \over H},
\end{equation}
and
\begin{equation}\label{eqn:znzbx}
B_x = 0.
\end{equation}
Field strength $B_0$ is such that the MRI wavelength is one quarter of
the domain height $H$.

In calculations with zero net magnetic flux centered at location~I,
field strength declines during the turbulent stage until the RMS
vertical MRI wavelength is less than two grid zones, and turbulence
ceases.  The true saturation level may correspond to a vertical MRI
wavelength less than the grid spacing.  The rest of the calculations
with zero net flux are carried out at location~II, where radiation
pressure is ten times gas pressure (section~\ref{sec:ic}).  To better
resolve the MRI wavelength at weak magnetic field strengths, a grid
consisting of $64\times 256\times 64$ zones is used except where
noted.  The parameters for the calculations with zero net magnetic
flux are listed in table~\ref{tab:znparms}, and some time- and
domain-averaged results are shown in table~\ref{tab:znresults}.
Initial pressures in table~\ref{tab:znparms} are written in terms of
the gas pressure at location~II.

\begin{deluxetable}{llccrclr}
\tablewidth{0pt}
\tablecaption{Parameters for simulations with zero net magnetic flux
\label{tab:znparms}}
\tablehead{Label&Name&$H/\lambda_{z,0}$&$\sigma/\ses$&
$P_0:$&$B_0^2/8\pi$&$:p_0$ & Duration\\ &&&&&&&/orbits}
\startdata
NS4  & Same total pressure&4  & \nodata &$  0 :$&$0.0160$&$:11$ &100 \\
NS4g & Same gas pressure  &4  & \nodata &$  0 :$&$0.0160$&$:1 $ &198 \\
RS4h & High opacity       &4  & 100	&$ 10 :$&$0.0160$&$:1 $ & 87 \\
RS4  & Radiation          &4  & 1	&$ 10 :$&$0.0160$&$:1 $ &140 \\
\enddata
\end{deluxetable}

\begin{deluxetable}{lccccccc}
\tabletypesize{\footnotesize}
\tablewidth{0pt}
\tablecolumns{7}
\tablecaption{Results from simulations with zero net magnetic
flux\label{tab:znresults}}
\tablehead{
\colhead{Label}
 & \colhead{Coupling}
 & \colhead{Field geometry}
 & \colhead{Total stress}
 & \colhead{Density range}
 & \colhead{Compression}
 & \colhead{Artificial} \\
\colhead{}
 & \colhead{$\sqrt{}\langle\langle B_z^2\rangle\rangle$/}
 & \colhead{$\langle\langle -B_xB_y/4\pi\rangle$}
 & \colhead{$10^3\langle\langle w_{xy}\rangle\rangle$}
 & \colhead{$\langle\langle\rho_{MAX}$}
 & \colhead{heating}
 & \colhead{viscous heating} \\
\colhead{}
 & \colhead{$(2c\Omega_0/3\chi)$}
 & \colhead{$/\langle B_z^2/8\pi\rangle\rangle$}
 & \colhead{$/(P_0+p_0)$}
 & \colhead{$/\rho_{MIN}\rangle\rangle$}
 & \colhead{$\langle\langle C\rangle\rangle / \langle\langle W\rangle\rangle$}
 & \colhead{$\langle\langle A\rangle\rangle
            / \langle\langle W\rangle\rangle$}\\
\colhead{(1)} & \colhead{(2)} & \colhead{(3)} &
\colhead{(4)} & \colhead{(5)} & \colhead{(6)} & \colhead{(7)}
}
\startdata
NS4	&\nodata &13.4&4.90&1.37&$-0.107 $&0.137 \\
NS4g	&\nodata &19.0&1.25&1.71&$-0.0272$&0.184 \\
RS4h	&81.0	 &12.4&6.49&1.52&$-0.0404$&0.131 \\
RS4	&0.214	 &16.2&2.08&2.04&$+0.208 $&0.0538\\
\enddata
\tablecomments{Results are averaged between 30 and 80 orbits.}
\end{deluxetable}

\subsection{Magnetic Field and Accretion Stress
\label{sec:znsaturation}}

We first examine how field strength in the turbulence depends on gas
pressure, in the absence of radiation.  The results are then compared
against similar calculations including coupling to photons.

For compressible flows, the gas pressure gradient term in the equation
of motion~2 can affect the development of the magnetic field.  Linear
perturbations have a compressive part if the background field includes
an azimuthal component \citep{bb94,ko00}.  Pressure effects may be
present also in well-developed turbulence.  In local shearing-box
calculations with zero net magnetic flux, the accretion stress at
saturation is proportional to the fourth root of the gas pressure
(T. Sano, in preparation).

\begin{figure}
\epsscale{0.65}
\plotone{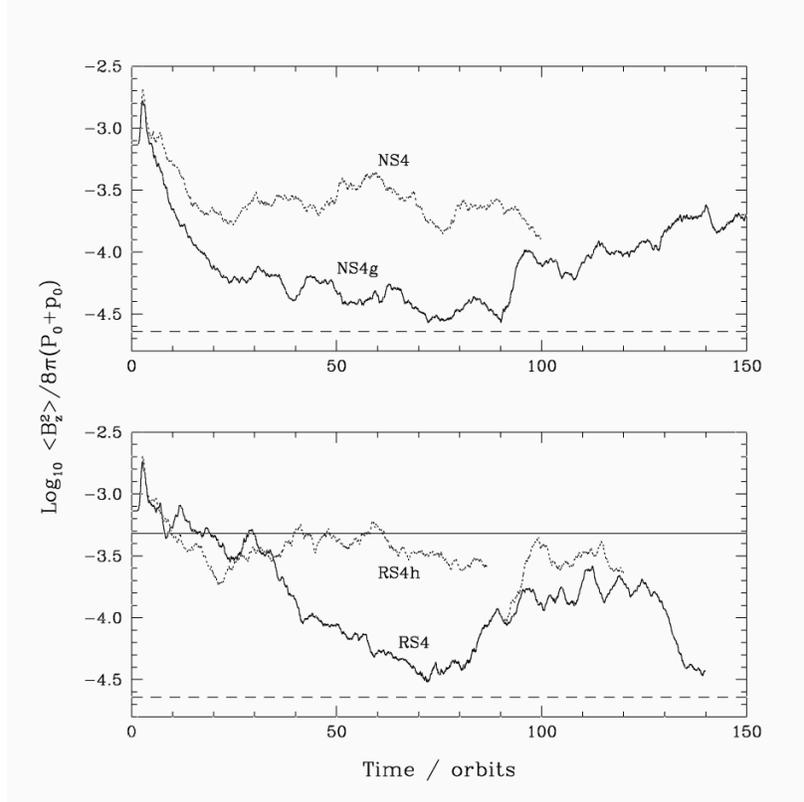}
\caption{Time variation of the domain-averaged pressure in the
vertical component of the magnetic field, in calculations with zero
net magnetic flux.  The magnetic pressure is measured in units of the
gas plus radiation pressure at location~II.  Results without radiation
are shown in the top panel.  The upper curve is from the run NS4 with
radiation pressure replaced by additional gas pressure, the lower from
the version NS4g with no extra pressure.  Results from runs including
radiation effects are in the lower panel.  The standard radiation run
RS4 is shown by the solid curve, the high-opacity version RS4h by the
dotted curve.  Beginning at 90 orbits, a second dotted curve marks a
version of the standard radiation run, with scattering opacity
suddenly increased to the value used in RS4h.  The horizontal solid
line indicates the pressure for which the vertical MRI wavelength is
equal to the distance photons diffuse per orbit, at the mean density
and standard opacity.  Dashed horizontal lines in both panels indicate
the pressure for which the vertical MRI wavelength is two grid zones.
\label{fig:znpmagz}}
\end{figure}

To compare directly with the radiation results described next, we
performed two calculations without radiation, using different initial
gas pressures.  In run NS4, the radiation pressure at location~II is
replaced by an equal amount of additional gas pressure.  In run NS4g,
no extra gas pressure is added, and the initial pressure is 11~times
less.  Results are shown in figure~\ref{fig:znpmagz}, upper panel.
The total accretion stress, averaged from 30 to 80 orbits, is 0.00490
in the higher-pressure calculation NS4, and 0.00124 in the
lower-pressure calculation NS4g.  The stresses are measured in units
of the initial gas plus radiation pressure at location~II.  In both
cases, the magnetic part of the stress is about three times the
hydrodynamic part.  Gas pressure increases over time in both runs,
owing to dissipation of the turbulence.  The mean gas pressures over
the same period are 1.08 and 0.209, in the same units.  During this
interval, run NS4 has mean gas pressure 5.2 times greater than NS4g,
and mean stress 4.0 times greater.  The relationship between stress
and pressure is steeper than $p^{1/4}$, perhaps due to the small
number of grid zones per vertical MRI wavelength in run NS4g.
However, the differences between these calculations with identical
initial magnetic fields suggest that stress does depend on pressure.
More evidence regarding the dependence can be obtained from the time
variation in the lower-pressure calculation, NS4g.  The accretion
stress increases beginning at 90~orbits.  The average between 130 and
180 orbits is 0.00260, 2.1 times greater than between 30 and 80
orbits.  The mean gas pressure is 0.364, 1.7 times greater than during
the earlier period.  All of these results indicate that the saturation
level of the magnetic field depends on gas pressure, when both net
magnetic flux and radiation effects are absent.

The effects on field strength of the coupling between radiation and
gas may be seen from the lower panel of figure~\ref{fig:znpmagz}.  In
the high-opacity calculation RS4h, the scattering opacity is 100 times
the electron scattering value, the vertical MRI wavelength is
initially 20~times the diffusion scale, and the photons are thoroughly
coupled to the flow throughout the run.  The field strength is similar
to that in run NS4, where radiation is replaced by extra gas pressure.
By contrast in the standard radiation calculation RS4, the scattering
opacity is set to the usual electron scattering value.  Initially the
vertical MRI wavelength is about twice the distance radiation diffuses
per orbit, and photons are moderately coupled to the gas.  The
domain-mean total magnetic pressure is always greater than initially
due to stretching along the azimuthal direction.  However the mean
pressure in the vertical component of the field declines over the
first ten orbits of turbulence, and thereafter remains less than its
initial value.  From 30 orbits onwards, the vertical MRI wavelength is
less than the diffusion scale, and photons are weakly coupled to
disturbances.  The magnetic field is weaker than in the high-opacity
case.  Averaged from 30 to 80 orbits, the total stress is $0.00208$ in
RS4, and $0.00649$ in the high-opacity version RS4h.  Field strength
in the standard-opacity run RS4 is comparable to that in the
low-gas-pressure run NS4g without radiation.  The two calculations
differ in that gas pressure rises more slowly with time in RS4.
Heating of the gas is almost entirely offset by the net emission of
radiation required to maintain thermal equilibrium.  In run NS4g,
radiation effects are not included, and the gas has no means of
cooling.

To examine possible effects of the choice of initial magnetic field,
we have also carried out a high-opacity run initialized with the
density, pressure, and magnetic field distributions from the
standard-opacity calculation RS4 at 90~orbits.  Over ten orbits, the
field grows in strength until it is similar to that in the
high-opacity calculation RS4h.  The total stress, averaged from 10 to
30 orbits after the increase in opacity, is 0.00525.

In these five calculations, the MRI generates fluctuations in magnetic
fields which lead to fluctuations in gas and radiation pressures.  If
radiation diffuses quickly, the radiation pressure is almost spatially
uniform, and gradients are smaller than those in gas pressure.  Gas
pressure gradients alone resist magnetic fluctuations.  On the other
hand, when radiation diffuses slowly, the spatial distributions of gas
and radiation are quite similar.  The radiation pressure gradient
works in the same way as the gas pressure gradient.  In this case, the
effective thermal pressure is the sum of the gas and radiation
pressures.  The saturation amplitude is found to be larger in the
calculations with larger effective pressure.

\begin{figure}
\epsscale{0.45}
\plotone{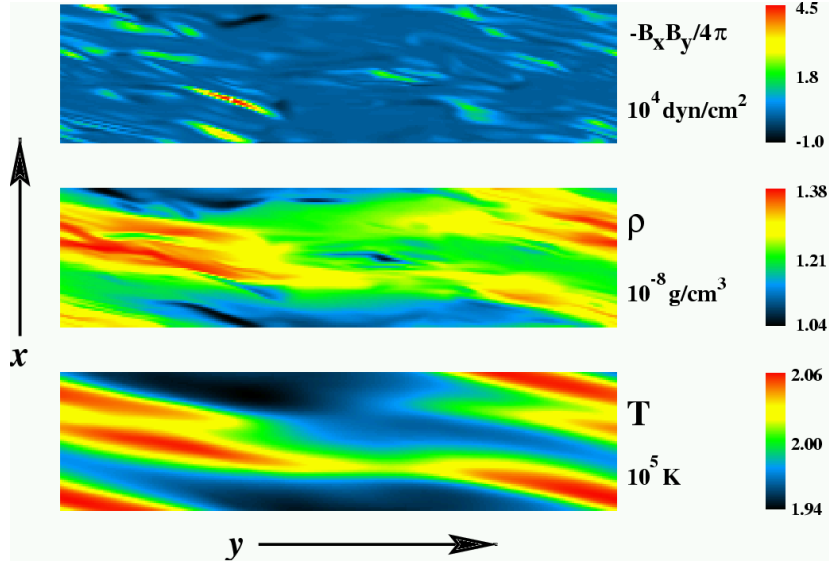}
\caption{Results from the standard calculation with radiation and zero
net magnetic flux.  Each panel is a slice through the midplane at
80~orbits.  Radius increases upwards, distance along the orbit to the
right.  Magnetic stress is shown at top, density in the middle panel,
and radiation temperature below.  The color scales are linear.  The
distance photons diffuse per orbit at the mean density is $1/7$ the
radial extent of the domain.
\label{fig:znimage}}
\end{figure}

The magnetic field is patchy in the standard radiation run RS4
(figure~\ref{fig:znimage}, top panel).  The vertical extent of the
patches is typically $1/10$ the domain height $H$, similar to the RMS
vertical MRI wavelength.  Domain-mean magnetic pressure also varies
over time, by almost an order of magnitude.  The vertical magnetic
pressure at 70~orbits is less than the initial total pressure by a
factor $28\,000$.  At 120 orbits, the ratio is only $5\,200$.  This
variation is probably not related to a long-term increase in gas or
radiation pressure.  Mean radiation pressure rises over 140~orbits by
only 35\%, gas pressure by 8\%.  The variation in stress is also
probably not due to changes in the degree to which radiation is tied
to the gas.  After the initial decline in field strength, the RMS
vertical MRI wavelength ranges from 0.25 to 0.74 times the distance
photons diffuse in an orbit, and coupling is weak.  The variation is
most likely related to the limited numerical resolution.  When the
magnetic field is at its strongest in calculation RS4, the RMS
vertical MRI wavelength is about ten grid zones.  However, when the
field is at its weakest, the wavelength is less than three zones.  In
a version of run RS4 using $32\times 128\times 32$ zones, the magnetic
field decays exponentially and turbulence ends, after the vertical MRI
wavelength falls below two grid zones.  In corresponding
low-resolution versions of the high-opacity and same-total-pressure
runs RS4h and NS4 with $32\times 128\times 32$ zones, magnetic fields
are stronger, the fastest-growing wavelengths are marginally resolved,
and turbulence is long-lasting.  The accretion stresses between 30 and
80 orbits, $0.00614$ and $0.00428$ respectively, are similar to those
in the higher-resolution versions.

The ratio of the magnetic stress to the vertical magnetic pressure in
the standard radiation calculation RS4, averaged from 30 to 80~orbits,
is 16.2.  The ratio varies with the strength of the field.  During
periods when the field is greatest, the ratio is about 12, and lies in
the range found in the calculations with net magnetic flux
(section~\ref{sec:bz}).  The field may be sufficiently well-resolved
in these periods.  At times when the RMS vertical MRI wavelength is
just longer than two grid zones, the stress is about 30 times the
vertical magnetic pressure.  The magnetic field is more nearly
azimuthal when it is weaker.  The field geometry varies with magnetic
pressure in a similar way in the calculations with high opacity RS4h
and without radiation NS4 and NS4g, within the range of magnetic
pressure the calculations have in common.  In the lower-resolution
versions of RS4h and NS4, the vertical magnetic pressure is about
three times less at the same magnetic stress.  This effect may be due
to marginal spatial resolution.

We conclude that the accretion stress depends on gas pressure in the
absence of radiation effects.  If radiation dominates total pressure,
the stress depends on both gas pressure and opacity.  When coupling
between gas and radiation is strong, radiation pressure plays a role
similar to extra gas pressure, and the stress is large.  When photons
are decoupled from turbulent motions, and compression is resisted
largely by gas pressure, the stress is less.

\subsection{Turbulent Fluctuations
\label{sec:znspatial}}

The gas is nearly isothermal on small scales in the standard radiation
run RS4, despite small-scale density fluctuations
(figure~\ref{fig:znimage}, middle and bottom panels).  Photons readily
diffuse between the regions of compression and expansion, so that the
radiation temperature is close to uniform.  Gas and radiation
temperatures differ by less than 1\% because the timescale for thermal
equilibration is much less than the orbital period.  In the
high-opacity calculation RS4h, the vertical MRI wavelength is longer
than the diffusion scale, the effective equation of state is that of a
$\gamma=4/3$ gas, and temperature fluctuations are comparable in
magnitude to density fluctuations.  The calculations NS4 and NS4g
without radiation show the expected adiabatic relation between
temperature and density.

\begin{figure}
\epsscale{0.65}
\plotone{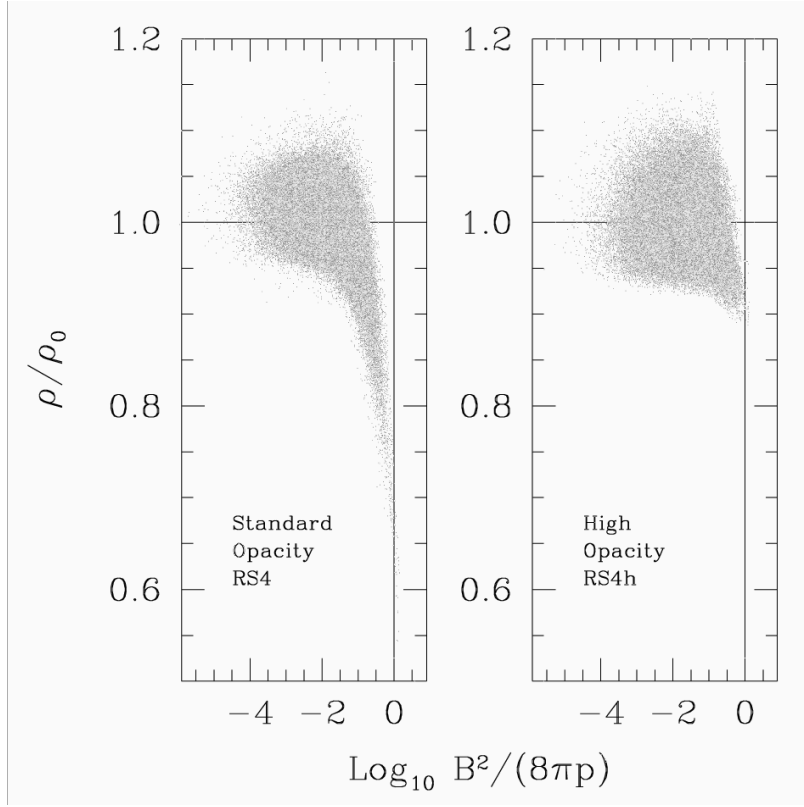}
\caption{Density versus the ratio of magnetic to gas pressure, in the
standard radiation calculation RS4 at 100~orbits (left), and the
high-opacity version RS4h at 80 orbits (right).  Every eleventh grid
zone is plotted.  Horizontal lines indicate the mean density.
Vertical lines mark equality of gas and magnetic pressures.  In the
standard-opacity run, lower densities are found in zones where the
magnetic pressure approaches the gas pressure.  In the high-opacity
run, radiation pressure provides additional support against
compression.  Zones having comparable gas and magnetic pressures show
only slightly reduced densities.  The ratio of the mean radiation
pressure to the mean gas pressure is 12.1~in the standard opacity run,
and 11.2 in the high-opacity run at these times.
\label{fig:znpmagvsp}}
\end{figure}

In the standard radiation run RS4, the density is lowest in regions
having magnetic pressure greater than gas pressure.  Where the
magnetic field is weak, the density is near its mean value
(figure~\ref{fig:znpmagvsp}).  The situation is different in the
high-opacity run RS4h, where radiation pressure resists compression
even in regions with magnetic pressure exceeding gas pressure.  The
mean density contrast in RS4h is 1.52, whereas that in RS4 is 2.04
despite the weaker magnetic field (table~\ref{tab:bzresults}).  In the
calculation NS4g with radiation removed, the mean density range
between 30 and 80~orbits is 1.71, intermediate between those in RS4
and RS4h.  The density contrast may be less than in RS4 because gas
pressure doubles during the first 30~orbits in NS4g, owing to
dissipation of the turbulence without net emission of photons.  In the
calculation NS4 with radiation replaced by extra gas pressure, the gas
pressure is greater than the magnetic pressure in every zone, and the
mean density contrast~1.37 is small.  In summary, density fluctuations
in the calculations with zero net magnetic flux are largest when
radiation is weakly coupled to the disturbances, and magnetic pressure
is greater than gas pressure.

\subsection{Heating
\label{sec:znheating}}

Of the four calculations with zero magnetic flux listed in
table~\ref{tab:znparms}, only one shows sustained net compression
heating.  In the standard radiation run RS4 between 30 and 80 orbits,
the total compression heating is equal to 21\% of the energy added to
the domain via accretion stresses (table~\ref{tab:znresults}).  In the
remainder of the simulations with zero magnetic flux, $P dV$ heating
is ineffective due to the absence or slow rate of radiation diffusion.
Compressive motions are instead damped by artificial viscosity.  In
the run NS4g with radiation removed, 18\% of the energy released by
accretion stresses is dissipated through artificial viscous heating.
Artificial viscosity is about as important in NS4g as compression
heating in RS4.  The artificial viscous heating in runs RS4h, NS4, and
NS4g is partly balanced by net expansion cooling, indicating pressures
are greater on average during expansion than during compression.  Of
the four calculations, only the standard radiation run RS4 shows
substantial heating by a physical mechanism.

%%%%%%%%%%%%%%%%%%%%%%%%%%%%%%%%%%%%%%%%%%%%%%%%%%%%%%%%%%%%%%%%%%%%%%%%%%%%%%%
\section{DISCUSSION
\label{sec:discussion}}

The variation of stress with opacity, the large density fluctuations,
and the radiation damping described in sections~\ref{sec:bz}
and~\ref{sec:zn} are all related to the amount of coupling between gas
and photons.  When the vertical MRI wavelength is longer than the
diffusion scale, and the inequality in equation~\ref{eqn:coupling} is
satisfied, the effective pressure which resists compression by
magnetic forces is due to gas and radiation pressures together.  When
the MRI wavelength is shorter than the diffusion scale, the effective
pressure is due to gas alone.  In regions of accretion disks where
radiation pressure greatly exceeds gas pressure, the degree of
coupling may be a crucial factor in determining the properties of the
turbulence.

\begin{figure}
\epsscale{0.75}
\plotone{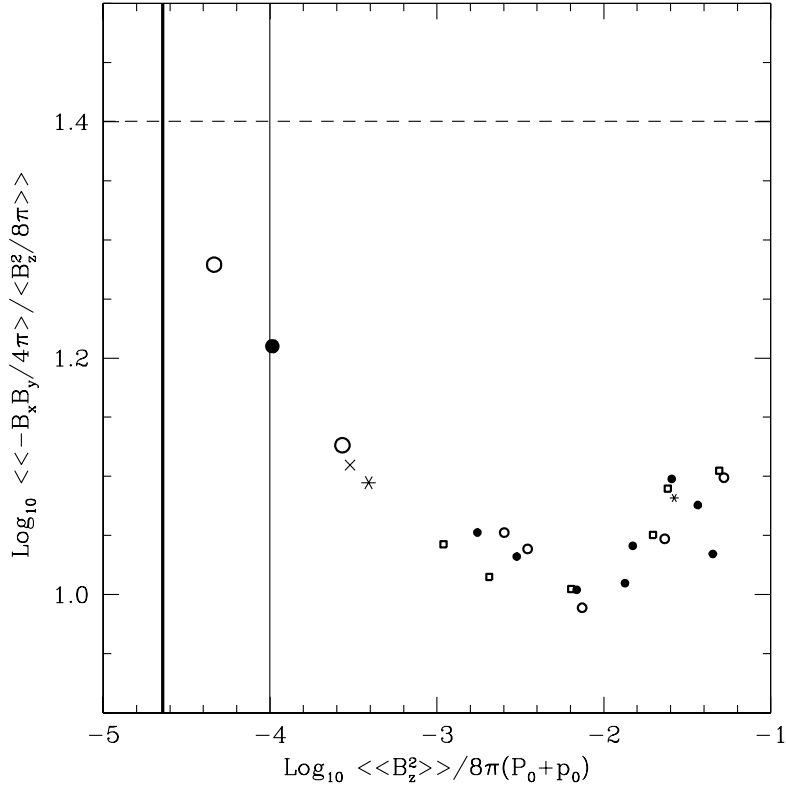}
\caption{Time-averaged ratio $s$ of magnetic stress to vertical
magnetic pressure, versus the vertical magnetic pressure, in all the
calculations listed in tables~\ref{tab:bzresults}
and~\ref{tab:znresults}.  Calculations having net magnetic flux are
shown by small symbols, those with zero net flux by large symbols.
Runs without radiation are indicated by open circles, those with high
opacity by stars, standard opacity by filled circles, and low opacity
by squares.  The continuation of the standard zero-net-flux run RS4
with increased opacity is shown by an `x'.  Vertical lines mark where
the RMS vertical MRI wavelength is equal to two grid zones in
calculations having the domain height divided into 32 (thin line) and
64~zones (thick).  A horizontal dashed line indicates the
stress-to-pressure ratio $s=8\pi$ for which the vertical MRI
wavelength may be equal to the distance photons diffuse per orbit, in
radiation-supported Shakura-Sunyaev disks.
\label{fig:coupling}}
\end{figure}

The orientation of the magnetic field may be measured by the ratio $s$
of the magnetic stress $\langle -B_xB_y/4\pi\rangle$ to the vertical
magnetic pressure $\langle B_z^2/8\pi\rangle$.  This ratio is similar
in calculations with and without radiation, and with gas--photon
coupling good or marginal.  The time-averaged values of $s$ in the
calculations listed in tables~\ref{tab:bzresults}
and~\ref{tab:znresults} are plotted in figure~\ref{fig:coupling}.  In
stratified isothermal simulations by \cite{ms00}, similar ratios of
magnetic stress to vertical magnetic pressure were found within two
scale heights of the midplane.  The ratio in each of these cases is
between eight and twenty.  The characteristic vertical scale of
turbulent fluctuations in these calculations is the RMS vertical MRI
wavelength $\langle\lambda_z^2\rangle^{1/2}=2\pi\langle
v_{Az}^2\rangle^{1/2}/\Omega$.  Owing to the small range of field
orientations $s$, the characteristic vertical scale may be
approximately determined from the accretion stress and gas density.

If magnetic fields in radiation-supported Shakura-Sunyaev disks have
orientations similar to those observed here, then the vertical MRI
wavelength $\lambda_z$ may be a fixed fraction of the distance photons
diffuse per orbit $l_D$.  From equation~\ref{eqn:coupling}, the
squared coupling ratio is
\begin{equation}
\left({\lambda_z\over l_D}\right)^2 = {3\chi\langle B_z^2\rangle\over
2c\Omega}.
\end{equation}
When a disk is in inflow equilibrium and its $R-\phi$ stress is
primarily due to magnetic forces, conservation of angular momentum
requires
\begin{equation}\label{eqn:amcons}
\langle -B_x B_y\rangle H = \dot M \Omega R_T,
\end{equation}
where $R_T$ includes relativistic corrections and the effect of the
overall flux of angular momentum through the disk.  If the MRI-driven
turbulence leads to magnetic fields with fixed orientation $s$, the
angular momentum conservation equation~\ref{eqn:amcons} constrains the
vertical magnetic pressure, and hence the squared coupling ratio
\begin{equation}
\left({\lambda_z\over l_D}\right)^2 = {3\chi\dot M\over s c H}R_T.
\end{equation}
Thermal equilibrium together with vertical hydrostatic balance between
gravity and radiation pressure determines the disk half-thickness $H$
in terms of the accretion rate.  Using this relation yields
\begin{equation}
{\lambda_z\over l_D} = \left({8 \pi \over s} {R_T R_z \over
R_R}\right)^{1/2},
\end{equation}
where $R_R$ is the correction factor that is the energy conservation
analog of $R_T$, and $R_z$ is the relativistic correction to the
vertical gravity.  For field orientation $s=8$ and unit correction
factors, the MRI wavelength is $1.8$ times the diffusion scale.  For
$s=20$, the ratio is $1.1$.

Thus, when a disk is in hydrostatic, thermal, and inflow equilibrium,
and its vertical support is primarily radiative, we can expect the
photons to be marginally coupled to the MHD turbulence in the
interior.  This result holds whether the stress scales with total
pressure, gas pressure alone, or some combination, as the stress law
is not specified in the derivation.  However, if an uneven dissipation
distribution makes the disk thinner than in the simplest picture of
complete radiation support \citep{sz94}, the coupling near the
midplane may be better than indicated here.  Given the strong
connection between coupling quality and compressibility that we have
seen, and the further likely connection between compressibility and
magnetic dissipation, the quality of coupling may be self-regulated in
some fashion.

\begin{figure}
\epsscale{0.75}
\plotone{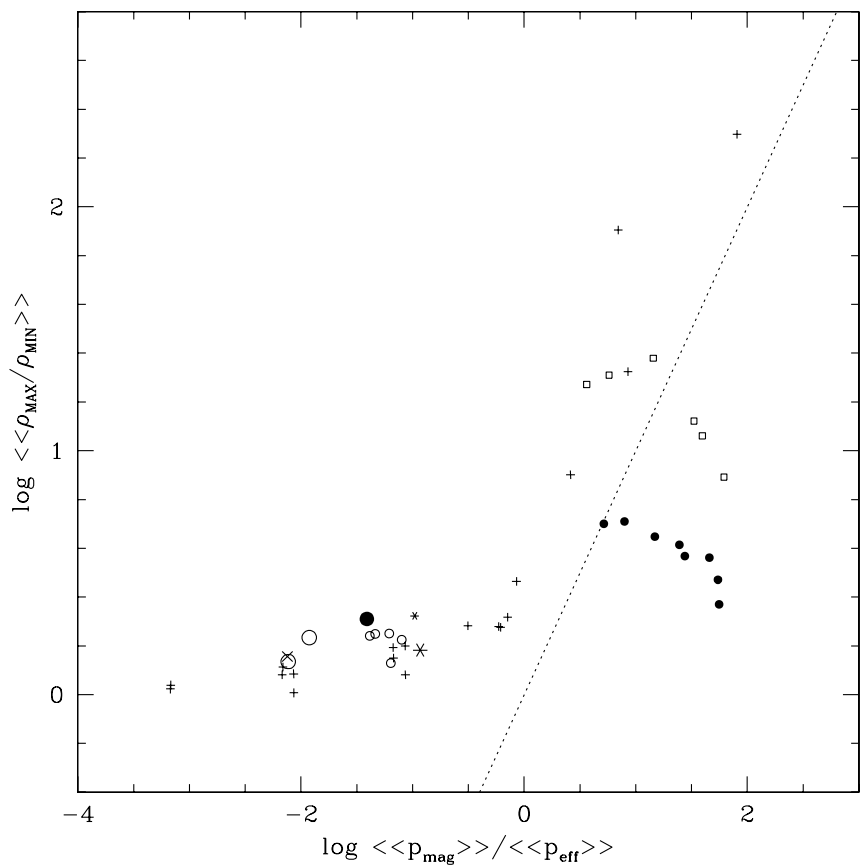}
\caption{Time-averaged density contrast versus the ratio of magnetic
to effective pressure in the calculations listed in
tables~\ref{tab:bzparms} and~\ref{tab:znparms}.  Results of
calculations discussed by TSS are shown also by plus signs.  Other
symbols are as in figure~\ref{fig:coupling}.  Effective pressure is
set equal to gas pressure except in calculations with high opacity,
where it is set to the sum of gas and radiation pressures.  A dotted
line marks $y=x$.
\label{fig:dcontrast}}
\end{figure}

The effects of diffusion on density contrasts in the turbulence are
summarized in figure~\ref{fig:dcontrast}.  Contrasts are large in
calculations with magnetic pressure greater than effective pressure.
The standard-opacity radiation runs with net vertical flux, shown by
small filled circles, are plotted using gas pressure for effective
pressure in all cases.  However, in those with the stronger magnetic
fields, longer MRI wavelengths mean that radiation is more tightly
tied to the turbulence.  The correct effective pressure is
intermediate between the gas and total pressures.  If effective
pressure is as large as total pressure, the case with strongest
magnetic field is shifted 2.9~decades to the left.  Cases with weaker
fields have weaker gas--radiation coupling, and are to be shifted
smaller distances to the left.  Also plotted are results from
calculations by TSS, listed in their table~2.  Together, the data
indicate density contrasts can be as great as the ratio of magnetic to
gas pressure provided radiation is not well-coupled to gas.  In the
axisymmetric calculations by TSS, the magnetic field geometry in the
transient turbulence is controlled by the initial condition.  At a
fixed magnetic pressure, fields more nearly aligned with the azimuthal
direction result in fluctuations with shorter vertical wavelengths.
Weaker coupling of radiation to gas over these smaller scales means
smaller effective pressures, so that magnetic forces produce greater
compression.  In the three-dimensional calculations, the field
geometry is determined by the action of the turbulence.  The field is
found to be more inclined from the azimuthal than in the fiducial case
of TSS, resulting in smaller density ranges at similar magnetic
pressures.  Density contrasts greater than twenty are demonstrated
here in long-lasting, three-dimensional turbulence.

%%%%%%%%%%%%%%%%%%%%%%%%%%%%%%%%%%%%%%%%%%%%%%%%%%%%%%%%%%%%%%%%%%%%%%%%%%%%%%%
\section{SUMMARY AND CONCLUSIONS
\label{sec:conclusions}}

We carried out three-dimensional MHD calculations of local patches of
a radiation-dominated accretion disk.  The vertical component of
gravity was neglected, and periodic boundaries were used in the
vertical direction.  Magneto-rotational instability led to magnetized
turbulence in which Maxwell stresses transported angular momentum
outwards.  A number of basic differences between results with and
without radiation are caused by the diffusion of photons with respect
to the material.  When opacity is high enough that radiation is locked
to gas over the length and time scales of fluctuations in the
turbulence, the accretion stress, density contrast, and dissipation
differ little from those in the corresponding calculations with
radiation replaced by extra gas pressure.  However, when radiation
diffuses each orbit a distance that is comparable to the RMS vertical
MRI wavelength, radiation pressure is less effective in resisting
squeezing.  Larger density fluctuations occur, and a non-linear
version of the radiation damping mechanism outlined by \cite{ak98}
converts $P dV$ work into photon energy.

The accretion stress is found to depend on the vertical magnetic flux
present.  If the net vertical field is large enough so the
corresponding MRI wavelength is at least $1/16$ the domain height, the
time-averaged stresses are similar whether photons and gas are
strongly or marginally coupled.  If there is no net vertical magnetic
flux, the field strength depends on the opacity, and is greater when
gas and radiation are dynamically well-coupled.  Over a wide range of
field strengths, the ratio of accretion stress to vertical magnetic
pressure is near the value required in radiation-supported
Shakura-Sunyaev models to make the vertical MRI wavelength equal to
the distance $\sqrt{\alpha}H$ that photons diffuse per orbit.  These
results indicate photons may be marginally coupled to turbulent eddies
in Shakura-Sunyaev disks accreting via internal magnetic stresses.
Our work is not the first indicating magnetic stress may scale with
gas pressure alone, rather than total pressure.  However, previous
suggestions were made either on the basis of an {\it ad hoc} search
for ways to cure instabilities besetting conventional disk models
\citep{sc81,tl84} or on the basis of a limitation to the magnetic
field strength placed by buoyancy \citep{sr84,sc89}.  The new point
here is that the inability of the magnetic field strength to track the
total pressure is due to photon diffusion effects.

Density contrasts greater than an order of magnitude are observed in
cases where magnetic pressure exceeds gas pressure, and photons partly
decouple from the gas.  Such large density variations may alter the
bulk radiation transport rate, the effective optical depth, and the
spectrum emerging from the disk photosphere.  The density fluctuations
involve repeated compression and expansion of fluid elements.
Diffusion of photons from compressed regions converts up to one
quarter of the released gravitational energy to radiation energy.

In the simulations discussed here, the heating results in secular
increases in radiation pressure, as the disk surface is omitted and
there is no means for cooling the flow.  Stationary disk structures
might be found using calculations including radiation losses.  An
accurate balance between heating and cooling can be obtained in such
calculations only if total energy is conserved.  In unstratified
calculations using a standard internal energy scheme, the majority of
the released energy disappears through numerical losses of magnetic
field.  In calculations conserving total energy during the magnetic
substep, approximate overall energy conservation is observed.  Such a
partial total energy scheme may be adequate for use in calculations of
disk vertical structure.  It is likely that the rates at which disks
heat and cool depend on the disk thickness.  Calculations in which the
thickness is allowed to vary may be useful in addressing the question
of the thermal stability of radiation-dominated disks.

%%%%%%%%%%%%%%%%%%%%%%%%%%%%%%%%%%%%%%%%%%%%%%%%%%%%%%%%%%%%%%%%%%%%%%%%%%%%%%%
\begin{acknowledgments}
This work was supported by the United States Department of Energy
under grant DFG-0398-DP-00215.  N. J. T. and J. H. K. thank the
organisers of the 2002 Aspen workshop on ``Astrophysical Disks'' and
the staff of the Aspen Center for Physics for their hospitality.
\end{acknowledgments}

%%%%%%%%%%%%%%%%%%%%%%%%%%%%%%%%%%%%%%%%%%%%%%%%%%%%%%%%%%%%%%%%%%%%%%%%%%%%%%%

\end{document}